\documentclass[10pt]{article}
\pdfoutput=1
\usepackage[utf8x]{inputenc}
\usepackage{amssymb,amsmath,mathrsfs,enumerate}
\usepackage{rotate,multicol}
\usepackage{graphicx,wrapfig}
\usepackage{tikz,graphics,color,fullpage,float,epsf,caption,subcaption}

\usepackage[colorlinks=true,
  linkcolor=red,
  urlcolor=blue,
  citecolor=blue]{hyperref}
\usepackage{color}
\usepackage{soul}
\usepackage{cite}
\usepackage{verbatim}
\usepackage[symbol]{footmisc}
\long\def\rpl#1!!#2!!{\textcolor{red}{#1} \textcolor{blue}{#2}}

\def \order(#1){{\cal O} \left(#1 \right)}

\newmuskip\pFqskip
\mathchardef\pFcomma=\mathcode`, 

\newcommand*\pFq[5]{%
  \begingroup
  \begingroup\lccode`~=`,
    \lowercase{\endgroup\def~}{\pFcomma\mkern\pFqskip}%
  \mathcode`,=\string"8000
  {}_{#1}F_{#2}\left({#3}{#4};#5\right)%
  \endgroup
}

\evensidemargin=\oddsidemargin
\def\Eqn#1{Eq.\ (\ref{#1})}
\def\Eqs#1#2{Eqs.\ (\ref{#1}) and (\ref{#2})}
\allowdisplaybreaks

\begin{document}

\begin{flushright}
  IPMU21-0026
\end{flushright}

\begin{center}
  {\Large \bf Prospects of light charged scalars in a three Higgs doublet model with $Z_3$ symmetry } \\
  \vspace*{1cm} {\sf Manimala
    Chakraborti$^{a,}$\footnote[1]{mani.chakraborti@gmail.com},~Dipankar
    Das$^{b,}$\footnote[2]{d.das@iiti.ac.in},~Miguel
    Levy$^{c,}$\footnote[4]{miguelplevy@ist.utl.pt},~Samadrita 
    Mukherjee$^{d,}$\footnote[3]{samadritamukherjee657@gmail.com},~Ipsita
    Saha$^{e,}$\footnote[5]{ipsita.saha@ipmu.jp}} \\
  \vspace{10pt} {\small \em 
    $^a$Astrocent, Nicolaus Copernicus Astronomical Center of the Polish Academy of Sciences, ul.Rektorska 4, 00-614 Warsaw, Poland\\
    $^b$Department of Physics, Indian Institute of Technology(Indore), Khandwa Road, Simrol, 453 552 Indore, India \\
    $^c$Centro de F\'{i}sica Te\'{o}rica de Part\'{i}culas-CFTP and Departamento de F\'{i}sica, Instituto Superior T\'{e}cnico,\\
    Universidade de Lisboa, Av Rovisco Pais, 1, P-1049-001 Lisboa, Portugal\\
    $^d$School of Physical Sciences, Indian Association for the Cultivation of Science, Kolkata 700 032, India \\
    $^e$Kavli IPMU (WPI), UTIAS, University of Tokyo, Kashiwa, 277-8583, Japan}
  
  \normalsize
\end{center}

\renewcommand*{\thefootnote}{\arabic{footnote}}
\setcounter{footnote}{0} 
\begin{abstract}
  The stringent constraints from the direct searches for exotic scalars
  at the LHC as well as indirect bounds from flavor physics measurements have imposed
  severe restrictions on the parameter space of new physics models featuring extended Higgs sectors.
  In the Type-II 2HDM, this implies a lower bound on the charged Higgs masses of $\order(600~{\rm GeV})$.
In this work we analyze the phenomenology
of a Z3HDM in the alignment limit focusing on the impact of flavor physics constraints on its parameter space.
  We show that the couplings of the two charged Higgs bosons in this model feature an additional suppression
  factor compared to Type-II 2HDM. This gives rise to a significant relaxation of the flavor physics constraints
  in this model, allowing the charged Higgs masses to be as low as $\order(200~{\rm GeV})$.
  We also consider the constraints coming from precision electroweak observables
  and the observed diphoton decay rate of the 125 GeV Higgs boson 
  at the LHC. The bounds coming from the direct searches of nonstandard Higgs bosons at the LHC, particularly
  those from resonance searches in the ditau channel, prove to be very effective
  in constraining this scenario further.
\end{abstract}

\bigskip

\section{Introduction}
\label{s:intro}
Two of the major tasks
that will be undertaken in the upcoming runs of the LHC and beyond comprise of
the precise determination of the properties of the 125 GeV Higgs boson
as well as direct searches for additional scalar particles.
The remarkable consistency between the predictions from the standard model (SM)
and the experimental data from the LHC so far has posed
strong challenges for new physics (NP) scenarios beyond the
standard model (BSM). 
A complimentary pathway to explore NP is provided by the
low energy precision measurements in flavor physics.
Measurements from dedicated flavor physics experiments
like BELLE, BABAR and LHCb has so far been largely in agreement with the SM,
providing stringent constraints on most of the BSM scenarios.

Introduction of additional
Higgs doublets has been one of the most popular choices
for new physics extensions beyond the SM.
The most minimal choice, the two Higgs doublet model (2HDM)~\cite{Branco:2011iw, Bhattacharyya:2015nca},
has been discussed widely in the literature from both
theoretical and phenomenological points of view.
In the well-known alignment limit, the lightest CP-even
Higgs boson of 2HDM can be SM-like in its tree-level couplings
to the fermions and vector bosons and thus can serve as
the 125 GeV scalar observed at the LHC~\cite{Gunion:2002zf, Carena:2013ooa, Das:2015mwa, Dev:2014yca, Pilaftsis:2016erj, Bhattacharyya:2013rya}.
The additional (pseudo) scalar and charged Higgs
bosons can give rise to interesting signatures at the LHC as well
as at various flavor physics experiments.
Consistency with the strong constraints
from the LHC and flavor observables often pushes the charged Higgs boson mass
in 2HDM towards the heavier end of the spectrum. It has been observed that
a combination of flavor physics measurements can exclude
the charged Higgs masses below $\order(600~{\rm GeV})$
in 2HDM of Type-II~\cite{Misiak:2017bgg}, where up and down-type quarks obtain their
masses from two different Higgs doublets. This bound on the charged scalar
masses can be somewhat relaxed in Type-I 2HDM where a single Higgs
doublet is responsible for generating masses of the up and down type quarks\cite{Arbey:2017gmh}.
This is because in Type-I 2HDM all the fermionic
couplings of the charged scalar are proportional to $\cot\beta$, with $\tan\beta$ being the ratio of the two 
vacuum expectation values (VEVs), as conventionally defined in 2HDMs. Therefore, the constraints
on the nonstandard scalars can be easily evaded by choosing $\tan\beta \gg 1$.
In this work we investigate the possibility of allowing lighter nonstandard
scalars without compromising the essential feature of Type-II 2HDM {\it i.e.}
two different doublets give masses to up and down quarks.

Moving beyond 2HDM, the most natural step ahead is to add one additional
Higgs doublet, leading to the three Higgs doublet model
(3HDM)~\cite{Branco:1980sz,Ivanov:2012fp, Ivanov:2012ry, Keus:2013hya, Camargo-Molina:2017klw,deMedeirosVarzielas:2019rrp}. As in the 2HDM case, it is possible to achieve an alignment limit corresponding
to a physical scalar resembling the properties of 125 GeV SM-like Higgs boson~\cite{Das:2019yad, Das:2014fea}.
In contrast to 2HDM, the scalar spectrum is much broader here,
offering a rich phenomenology in both high and low energy experiments.
Most importantly, the presence of additional nonstandard Higgs bosons
leads to significant modifications in the flavor changing neutral and charged
current processes compared to 2HDM. In Ref.~\cite{Das:2019yad}, it was shown
that the conditions for alignment limit in 3HDM
can be parametrized by a set of simple equations closely mimicking
those of the CP conserving 2HDM. Using the example
of $Z_3$-symmetric 3HDM (Z3HDM), it was observed that the analytic conditions
can be easily implemented in a realistic scenario, making way for efficient
numerical analysis. In the present work we explore the
phenomenological aspects of the alignment limit in Z3HDM
with an emphasis on the effects of flavor
physics constraints on its parameter space.
We show that the constraints on the parameter space stemming from the
interplay of various flavor physics data are
notably relaxed compared to those in the Type-II 2HDM.
Such a relaxation of constraints transpires from the presence of an additional suppression
in the couplings of the charged Higgs bosons in the model compared
to Type-II 2HDM.
We prescribe a simple analytical set up that automatically guarantees
agreement with the $\rho$-parameter constraints
as well as bounds arising from the measurement of Higgs to diphoton decay rate at the LHC.
We also study the effect of the bounds coming from
direct searches for additional Higgs bosons at the LHC on the parameter space
of out interest.

Our paper is organized as follows. 
In Sec.~\ref{s:model} we describe the scalar sector and Yukawa structure of Z3HDM.
The constraints from flavor physics observables are analyzed in Sec.~\ref{s:flavcons}.
We calculate the diphoton decay rate for this model in Sec.~\ref{s:diphoton}. The limits coming
from the direct searches at the LHC are discussed in Sec.~\ref{s:direct}. Finally, we
summarize our results in Sec.~\ref{s:summary}.



\section{The Model: 3HDM with $Z_3$ symmetry}
\label{s:model}


The study of nHDMs leads to a rich phenomenology, as well as a sharp increase of the number of parameters, due to the addition of a SM-like Yukawa structure for each doublet, in general.  
Thus, the diagonalization of the 
mass matrices will not lead to the simultaneous diagonalization of all the associated Yukawa matrices, which will bring in flavor changing neutral currents (FCNCs) at the tree-level mediated by the neutral scalars. Since experimental data suggest that FCNCs are highly suppressed~\cite{Zyla:2020zbs}, one interesting path to undertake 
is the study of models with natural flavor conservation (NFC)~\cite{Glashow:1976nt}. In these cases, each type of fermion is coupled to a single scalar doublet, ensuring the simultaneous diagonalization 
of the Yukawa and Mass matrices, leading to the absence of FCNCs at tree-level.

Within the framework of 2HDMs, there are four known types of models featuring NFC, which amount to the distinct possibilities of coupling each scalar to the fermions. Contrary to what one might 
expect, enlarging the framework to a 3HDM only adds one more nonequivalent possibility which ensures NFC.  The different types of models, characterized by their Yukawa structures, are shown in Table~\ref{tab:NFC}. In this work, we focus on the case unique to models with more than two Higgs doublets, sometimes referred to as democratic or type-Z 3HDM \cite{Cree:2011uy,Akeroyd:2016ssd,Logan:2020mdz}.
\begin{table}
\centering
\begin{tabular}{ c|cccc|c } 
 \hline
 \hline
  fermion & Type-I & Type-II & Type-X & Type-Y & Democratic \\ 
 \hline
 \hline
 $u$ & $\phi_3$ & $\phi_3$ & $\phi_3$ & $\phi_3$ & $\phi_3$  \\ 
 $d$ & $\phi_3$ & $\phi_2$ & $\phi_3$ & $\phi_2$ & $\phi_2$  \\ 
 $\ell$ & $\phi_3$ & $\phi_2$ & $\phi_2$ & $\phi_3$ & $\phi_1$ \\
 \hline
\end{tabular}
 \caption{\small All nonequivalent possibilities for models featuring NFC. The first four types can be realized within 2HDMs, while the last requires at least a 3HDM. 
}
 \label{tab:NFC}
\end{table}
However, such a democratic Yukawa structure can be implemented in more than one ways. Here, we make use of the matching number of fermionic and scalar doublets generations to endow the 3HDM with a $Z_3$
symmetry\footnote{It should be noted that a democratic 3HDM which features a similar Yukawa structure can also be obtained by imposing a $Z_2 \times Z_2$ symmetry \cite{Akeroyd:2016ssd,Logan:2020mdz}.}.
By doing so, we are able to find suitable charge assignments for both the fermions and the scalar doublets such that the NFC model ensues.

In our current setup, only scalars and right-handed fermionic fields may transform nontrivially 
under the $Z_3$ symmetry. Namely, we require
\begin{subequations}
	\label{e:transf}
	\begin{eqnarray}
	\label{eq:scalar-transf}
	\phi_1 \to \omega \, \phi_1 \,, \qquad & \phi_2 \to \omega^2 \phi_2 \,,  \\
	\ell_R \to \omega^2 \ell_R \,,  \qquad & n_R \to \omega \,n_R \,,  \label{eq:ferm-transf}
	\end{eqnarray}
\end{subequations}
where $\omega=e^{2 \pi i /3}$, and $n_R$ ($\ell_R$) are the right-handed down-type quarks (leptons), as to clearly distinguish between the flavor and mass eigenstates.  
By taking all other fields to transform trivially under the $Z_3$ symmetry, it becomes clear that $\phi_1$ couples to the leptons, whereas 
$\phi_2$ and $\phi_3$ couple to the down- and up-type quarks, respectively.  As such, we achieve a Yukawa structure which ensures NFC and the absence of FCNCs at tree-level. 


\subsection{Scalar sector}

While there are more than one ways to achieve a democratic Yukawa structure, the different choices will lead to different scalar potentials. The most general scalar potential for a Z3HDM
obeying the symmetry in \Eqn{eq:scalar-transf} is given by~\cite{Das:2019yad,Chakraborty:2019zas,Alves:2020brq,Boto2021}
\begin{eqnarray}
  V &=& m_{11}^2(\phi_1^\dagger \phi_1) +  m_{22}^2(\phi_2^\dagger 
  \phi_2) +  m_{33}^2 (\phi_3^\dagger \phi_3)  - \left(m_{12}^2(\phi_1^\dagger \phi_2) +  m_{23}^2(\phi_2^\dagger 
  \phi_3) +  m_{13}^2 (\phi_1^\dagger \phi_3) + {\rm h.c.} \right)   \nonumber \\
   &&+\lambda_1(\phi_1^\dagger 
  \phi_1)^2 + \lambda_2(\phi_2^\dagger \phi_2)^2 + 
  \lambda_3(\phi_3^\dagger \phi_3)^2  \nonumber \\ 
  && + \lambda_4 (\phi_1^\dagger \phi_1)( \phi_2^\dagger \phi_2) + \lambda_5 (\phi_1^\dagger \phi_1)( \phi_3^\dagger \phi_3) + 
  \lambda_6 (\phi_2^\dagger \phi_2)( \phi_3^\dagger \phi_3) \nonumber \\
  && + \lambda_7 (\phi_1^\dagger \phi_2)( \phi_2^\dagger \phi_1) +
  \lambda_8 (\phi_1^\dagger \phi_3)( \phi_3^\dagger \phi_1)+ \lambda_9
  (\phi_2^\dagger \phi_3)( \phi_3^\dagger \phi_2) \nonumber \\
  && + \left[\lambda_{10} (\phi_1^\dagger \phi_2)( \phi_1^\dagger \phi_3) 
    + \lambda_{11} (\phi_2^\dagger \phi_1)( \phi_2^\dagger \phi_3)+
    \lambda_{12} (\phi_3^\dagger \phi_1)( \phi_3^\dagger \phi_2) + {\rm h.c.} \right] \,,
  \label{e:potential}
\end{eqnarray}
where we allow the presence of the soft-breaking terms $m^2_{ij}$, $i\neq j$, since they will be of some importance for 
regulating the charged Higgs contribution to the diphoton decay amplitude\cite{Bhattacharyya:2014oka}.
For simplicity, we take all the parameters of the scalar potential to be real, so that the neutral scalars can be easily classified as CP-even and CP-odd bosons.

After the electroweak symmetry breaking (EWSB), the scalar doublets can be decomposed
in terms of the component fields as
\begin{eqnarray}
  \phi_k = \frac{1}{\sqrt{2}}
  \begin{pmatrix} \sqrt{2}\, w_k^+ \\  v_k + h_k 
    + i z_k  \end{pmatrix} \;, \quad k = 1,2,3,
  \label{e:scalvev}
\end{eqnarray}
where $v_k$ denotes the VEV of the field $\phi_k$ ($\left<\phi_k\right> = v_k/\sqrt{2}$). For notational convenience, the VEVs are expressed as
\begin{eqnarray}\label{eq:vevs}
  v_1 = v \cos\beta_1 \cos\beta_2 \;, \quad v_2 = v \sin\beta_1 \cos\beta_2 \;, \quad
  v_3 = v \sin\beta_2\;,
\end{eqnarray}
where $v = \sqrt{v_1^2+v_2^2+v_3^2}$ is the usual electroweak (EW) VEV.
The inclusion of three scalar doublets will give rise to four charged scalar particles, $H_{1,2}^\pm$, three CP-even neutral ones $h, H_{1,2}$, as well as two 
CP-odd neutral particles $A_{1,2}$, where the remaining fields are the usual Goldstone bosons $w^\pm, \zeta$. These physical particles can be obtained by rotating the fields 
onto the mass basis. For the charged and pseudoscalar sectors, we can obtain the physical scalars by performing the following $3\times3$ rotations,
\begin{equation}
  \begin{pmatrix} w^\pm\\ H_1^\pm \\ H_2^\pm\end{pmatrix} =
    \mathcal{O}_{\gamma_2}\mathcal{O}_\beta
    \begin{pmatrix} w_1^\pm\\ w_2^\pm \\ w_3^\pm\end{pmatrix}\; , \quad
      \quad
      \begin{pmatrix} \zeta\\ A_1 \\ A_2\end{pmatrix} =
        \mathcal{O}_{\gamma_1} \mathcal{O}_\beta 
        \begin{pmatrix} z_1\\ z_2 \\ z_3\end{pmatrix} ,
        \label{charged-and-CP-odd}
\end{equation}
where, the rotation matrices are defined as
\begin{eqnarray}\label{eq:gammaRots}
  {\cal O}_{\gamma_1} =
  \begin{pmatrix}
    1 & 0 & 0 \\
    0 & \cos\gamma_1 & -\sin\gamma_1 \\
    0 & \sin\gamma_1 & \cos\gamma_1 \end{pmatrix}  \label{e:Ogamma1} \,,
  \quad{\cal O}_{\gamma_2} =
  \begin{pmatrix}
    1 & 0 & 0 \\
    0 & \cos\gamma_2 & -\sin\gamma_2 \\
    0 & \sin\gamma_2 & \cos\gamma_2 \end{pmatrix}  \label{e:Ogamma2} \, ,
\end{eqnarray}
and
\begin{eqnarray}\label{eq:betaRot}
  {\cal O}_{\beta} =
  \begin{pmatrix} \cos\beta_2 \cos\beta_1 & \cos\beta_2 \sin\beta_1
    &  \sin\beta_2 \\
    -\sin\beta_1 & \cos\beta_1  &  0  \\
    -\cos\beta_1 \sin\beta_2 & -\sin\beta_1\sin\beta_2 & \cos\beta_2 
  \end{pmatrix}.
\end{eqnarray}
For the CP-even sector, we can obtain the physical mass basis through
\begin{eqnarray}
  \label{e:ab3hdm}
\begin{pmatrix}
    h \\
    H_1 \\
    H_2 
  \end{pmatrix}
  &=&{\cal O}_ \alpha 
  \begin{pmatrix}
    h_1 \\
    h_2 \\
    h_3 
  \end{pmatrix}
  \label{CP-even}
\end{eqnarray}
where
\begin{subequations}
  \label{e:Oa}
  \begin{eqnarray}
    {\cal O}_\alpha &=& {\cal R}_3 \cdot  {\cal R}_2\cdot {\cal R}_1 \,,
  \end{eqnarray}
with
  \begin{equation}
    \label{e:R}
          {\cal R}_1 = \begin{pmatrix}
                     \cos \alpha_1 & \sin \alpha_1 & 0 \\
            -\sin \alpha_1 & \cos \alpha_1 & 0 \\
            0 & 0 & 1 \end{pmatrix}
          \,, \quad {\cal R}_2 = \begin{pmatrix}
            \cos \alpha_2 & 0 & \sin \alpha_2  \\
            0 & 1 & 0 \\
            -\sin \alpha_2 & 0 & \cos \alpha_2 
          \end{pmatrix}\,,  \quad
          {\cal R}_3 = \begin{pmatrix}
            1 & 0 & 0 \\
            0 & \cos \alpha_3 &  \sin \alpha_3  \\
            0 & -\sin \alpha_3 & \cos \alpha_3 
          \end{pmatrix}.
  \end{equation}
\end{subequations}
For more details on the analysis of the scalar potential, we refer the reader to
Appendix~\ref{app:ScalarSector}.

The existence of nonstandard neutral CP-even scalars in the Z3HDM, in general, leads to a deviation 
of the couplings of the physical scalar $h$ from the respective SM predictions. However, the data obtained 
from the LHC runs shows a good agreement of the experimental data to the SM prediction for the Higgs 
signal strengths \cite{Sirunyan:2018koj, Aad:2019mbh}. This motivates us to work in
the alignment limit which is a set of conditions such that the 
lightest CP-even scalar mimics the SM-Higgs in its tree-level couplings, automatically respecting the agreement between the experimental data and the corresponding 
SM predictions for the Higgs signal strengths.
For our Z3HDM, the conditions for alignment are given by\cite{Das:2019yad}
\begin{eqnarray}
	\alpha_1=\beta_1 \,, \qquad
	\alpha_2=\beta_2 \,.
\end{eqnarray}
 As more data accumulate in the future runs of the High-Luminosity LHC (HL-LHC), 
the possibility of deviating from the alignment limit will become increasingly constrained, if no BSM signals are detected.

\subsection{Yukawa sector and charged Higgs couplings}
\label{s:yuk}
The quark Yukawa Lagrangian of the Z3HDM can be written as
\begin{equation}
  \mathcal{L} =  - Y_d  \overline{Q}_L \phi_2 n_R - Y_u \overline Q_L \widetilde{\phi}_3 p_R + {\rm h.c.} \,,
  \label{yukla1}
\end{equation}  
where $Q_L \equiv ( p_L, n_L )^T$ denotes the
$SU(2)_L$ left-handed quark doublet field, $p_R$ the up-type right-handed quarks, and $Y_{d,u}$ are the respective $3\times3$ Yukawa matrices in flavor space.
After EWSB, the mass matrices of the down and up-type quarks are given by
\begin{equation}
  M_d = Y_d \frac{v_2}{\sqrt{2}}\;; \quad M_u = Y_u \frac{v_3}{\sqrt{2}} \,.
\end{equation}
As usual, we can redefine the quark fields to rotate into the mass basis through
\begin{equation}
  d_L = \mathcal{D}_L\, n_L \,, \quad d_R = \mathcal{D}_R\, n_R\,, \quad u_L = \mathcal{U}_L\, p_L \,, \quad u_R = \mathcal{U}_R\, p_R \,,
\end{equation}
which, in turn, will diagonalize the mass matrices through the bi-unitary transformation
\begin{subequations}
	\begin{eqnarray}
	&& D_d = \mathcal{D}_L\, M_d\, \mathcal{D}_R^\dagger = {\rm diag}  (m_d,~ m_s,~ m_b) \,,\\
	&& D_u = \mathcal{U}_L\, M_u\, \mathcal{U}_R^\dagger = {\rm diag} (m_u,~ m_c,~ m_t) \,.
	\end{eqnarray}
\label{e:bidiag}	
\end{subequations}
Similar to the SM, the CKM matrix is defined as $V = \mathcal{U}_L \mathcal{D}_L^\dagger$.
As intended, our model does not have any FCNC at the tree-level, and the Higgs signal
strengths will also be compatible with the corresponding SM expectations in the alignment limit. However, the presence of charged scalars brings forth new channels for loop contributions to several 
flavor observables such as neutral meson oscillations and $b \to s \gamma$. In fact, 
these processes are quite restricted from experiments, and thus are usually used to place lower bounds on the 
nonstandard scalar masses, as their contributions must be kept in check. Thus, it becomes important to study the charged 
scalar couplings to the fermions, as these will govern the vertices responsible for these processes at the one-loop level.

Given its importance, we focus on the original quark Yukawa Lagrangian containing the charged Higgs couplings,
\begin{subequations}
	\begin{eqnarray}
	\mathcal{L}^Q_c &=& - Y_d\, \overline p_L\, n_R\, w_2^+ + Y_u^\dagger\, \overline p_R\, n_L\, w_3^+ + {\rm h.c.} \, \\
	&=& \frac{\sqrt{2}}{v} \overline u \left[ -\frac{1}{s_{\beta_1}c_{\beta_2}} w_2^+ (V \, D_d) P_R +
	\frac{1}{s_{\beta_2}} w_3^+ (D_u \, V) P_L \right] d + {\rm h.c.} \,,
	\label{yukla2}
	\end{eqnarray}
\end{subequations}
where, in the last step, we have rotated into the quark mass basis. Our goal is to arrive at couplings among  
the physical fields, as so we further rewrite the Lagrangian in the scalar mass basis. 
Using $X = \mathcal{O}_\beta^T \mathcal{O}_{\gamma_2}^T$, Eq.~\eqref{yukla2} becomes
%
	\begin{eqnarray}
	\mathcal{L}^Q_c &=& \frac{\sqrt{2}}{v} H_1^+ \overline u \left[ \frac{X_{32}}{s_{\beta_2}} (D_u \, V) P_L
	-\frac{X_{22}}{s_{\beta_1}c_{\beta_2}} (V \, D_d) P_R \right] d \nonumber \\
	&& +\frac{\sqrt{2}}{v} H_2^+ \overline u \left[ \frac{X_{33}}{s_{\beta_2}} (D_u \, V) P_L
	-\frac{X_{23}}{s_{\beta_1}c_{\beta_2}} (V \, D_d) P_R \right] d 
	+ {\rm h.c.} \,,  
	\label{yukla3}
	\end{eqnarray}
%
which describes the vertices between the physical charged scalars to the physical quarks. The same process can be repeated to 
obtain the leptonic couplings:
\begin{eqnarray}
  \mathcal{L}^\ell_c = -\frac{\sqrt{2}}{v} H_1^+\, \overline \nu \frac{X_{12}}{c_{\beta_1}c_{\beta_2}} D_\ell\, P_R\, \ell
  -\frac{\sqrt{2}}{v} H_2^+\, \overline \nu \frac{X_{13}}{c_{\beta_1}c_{\beta_2}} D_\ell\, P_R\, \ell
  + {\rm h.c.} \,,
  \label{yukla4}
\end{eqnarray}
where, $\ell \equiv (e,\mu,\tau)^T$, $\nu\equiv (\nu_e,\nu_\mu,\nu_\tau)^T$ and
$D_\ell = {\rm diag}(m_e,m_\mu, m_\tau)$.
In the following, we will focus mostly on the consequences of quark flavor observables. Hence, to better grasp the model's implications, 
it is helpful to substitute the $X_{ij}$ elements explicitly following Eqs.~\eqref{eq:gammaRots} and \eqref{eq:betaRot}, recasting the charged 
Higgs couplings to quarks as
\begin{subequations}
\label{e:yukh1h2}
	\begin{eqnarray}
	\mathcal{L}^Q_{H_1^\pm} &=& -\frac{\sqrt{2}}{v} H_1^+\, \overline u \left[  \cot{\beta_2} \sin{\gamma_2} (D_u \, V) P_L
	+ \tan\beta_2\left(\frac{\cot{\beta_1}\cos{\gamma_2}}{\sin{\beta_2}}+\sin{\gamma_2}\right) (V \, D_d) P_R \right] d
	+ {\rm h.c.} \,, 
	\label{yukla5} \\
	\mathcal{L}^Q_{H_2^\pm} &=&
	\frac{\sqrt{2}}{v} H_2^+\, \overline u \left[  \cot\beta_2\cos{\gamma_2} (D_u \, V) P_L
	- \tan\beta_2\left(\frac{\cot\beta_1 \sin\gamma_2}{\sin{\beta_2}}-\cos\gamma_2\right) (V \, D_d) P_R \right] d 
	+{\rm h.c.} 
	\label{yukla6}
	\end{eqnarray}
\end{subequations}

One noteworthy observation is the similarity between the Z3HDM and the type-II 2HDM. In fact, both are NFC models, where the 
difference lies in the fact that, in the Z3HDM, the lepton Yukawa couplings have a dedicated doublet, whereas in the type-II 2HDM 
the leptons share the doublet responsible for the down type quark masses. The resemblance
can be made more explicit by noting that  
due to the $Z_3$ charge assignments of the scalar doublets, $\phi_1$ is responsible for the lepton masses, which are generally 
much lower than the quark masses. Combining this with the relation between each individual VEV and the EWSB seen in 
Eq.~\eqref{eq:vevs}, it seems reasonable to assume $v_1 \ll v_2, v_3$, which is achieved by taking large values of $\tan\beta_1$, while still remaining 
in a perturbative regime for the $\tau$-Yukawa coupling. In this regime, where $\cot\beta_1 \ll 1$, the $\cot\beta_1$ dependency of the 
charged Higgs couplings of Eqs.~\eqref{yukla5} and \eqref{yukla6} can be neglected, and the couplings become similar to those of 
the type-II 2HDM, relaxed by either $\cos\gamma_2$ or $\sin\gamma_2$, which are always less than
one. Indeed, by comparing with the corresponding
couplings in the type-II 2HDM \cite{Branco:2011iw}, 
\begin{equation} \label{eq:2hdmcharged}
\mathcal{L}^{\text{2HDM-II}}_{H^\pm} = \frac{\sqrt{2} H^+}{v} \big[ \cot\beta\, \overline{u}_R \left( D_u \, V\right) d_L + \tan\beta\, \overline{u}_L \left( V \, D_d\right) d_R \big]+
{\rm h.c.} \,,
\end{equation}
we can identify $\tan\beta$ of 2HDM-II with $\tan\beta_2$ of Z3HDM, since both control the ratio $v_u/v_d$, where $v_{u\,(d)}$ 
are the VEVs of the scalars that couple to the up (down) quarks, respectively.
If we further consider a scenario where either $H_1^\pm$ or $H_2^\pm$ is relatively heavy ( $\gtrsim 5$ TeV), while keeping the other 
relatively light ($\lesssim 1$ TeV), then the heavy particle decouples and its contribution will be negligible, and our effective theory 
becomes similar to a type-II 2HDM scenario. The striking difference is that while one of the scalars is decoupled, the effective theory still 
retains some consequences of the full theory. 
In order to exemplify, we consider a scenario where $H_2^\pm$ is decoupled and $\tan\beta_1\gg 1$.
In this case the $H_1^\pm$ couplings of \Eqn{yukla5} can be approximated as:
\begin{equation}\label{eq:z3hdmlimit}
\mathcal{L}^{\text{Z3HDM}}_{H_1^\pm} \approx -\sin\gamma_2\cdot \frac{\sqrt{2} H_1^+}{v}  \big[ \cot\beta_2\, \overline{u}_R 
\left( D_u \, V\right) d_L + \tan\beta_2\, \overline{u}_L \left( V \, D_d\right) d_R \big] +{\rm h.c.}
\end{equation}
Comparing with \Eqn{eq:2hdmcharged}, we notice the remarkable similarity with the type-II 2HDM
except for the fact that the couplings are reduced in strength by a factor of $\sin\gamma_2$.
This will play an important role in diluting the constraints from flavor data compared
to those in the type-II 2HDM, which we will discuss in the next section.

\section{Constraints from flavor data}
\label{s:flavcons}
Since compliance with flavor data is continuously pushing the lower bound on the mass of the charged Higgs of the type-II 2HDM upwards, the 
relaxation due to $\gamma_2$ in this effective 2HDM can easily justify lower masses for new charged particles, while still remaining within the 
experimental limits for the new physics (NP) contributions to the flavor processes. 

In order to make the discussion concrete, we analyze the resulting bounds coming from flavor data. We restrict ourselves to the analysis of the 
NP contributions to the radiative decay $b \to s \gamma$, as well as the bounds coming from the $B$ meson oscillations, $\Delta M_{B_{s,d}}$\footnote{The constraints from $\Delta M_K$ are
much weaker.}.  
 We make use of the FlavorKit \cite{Porod:2014xia} functionalities within SPheno \cite{Porod:2003um, Porod:2011nf}, compiled by SARAH 
 \cite{Staub:2013tta}, explicitly retaining contributions up to one-loop only. In order to gain some qualitative insights into the processes and 
 phenomenologies at hand, we refer the reader to appendix \ref{app:flavorObs}, where we provide analytic expressions for the relevant processes.  
 It is, however, easy to note that in models with no tree-level FCNCs, the only one-loop NP contributions to both $b \to s \gamma$ as well as 
 $\Delta M_{B_{s,d}}$ will come from the charged Higgs couplings. 
 Therefore, these observables will be governed by a set of five parameters, namely, $(\tan\beta_1,\, \tan\beta_2,\,  \gamma_2,\, 
  m_{H_1^+},\,  m_{H_2^+})$.

As we mentioned earlier, the Z3HDM where one charged Higgs is decoupled from EW scale dynamics becomes a relaxed type-II 2HDM 
effective scenario. Namely, a remnant of the full theory survives as a damping of the usual type-II 2HDM charged scalar couplings, which will 
in turn result in a relaxation of the bounds that are found for the type-II 2HDM. As such, we initially focus on this case where 
one of the charged scalars is decoupled, featuring the relaxation of the bounds.
\begin{figure}
\centering
\includegraphics[width=0.9\textwidth]{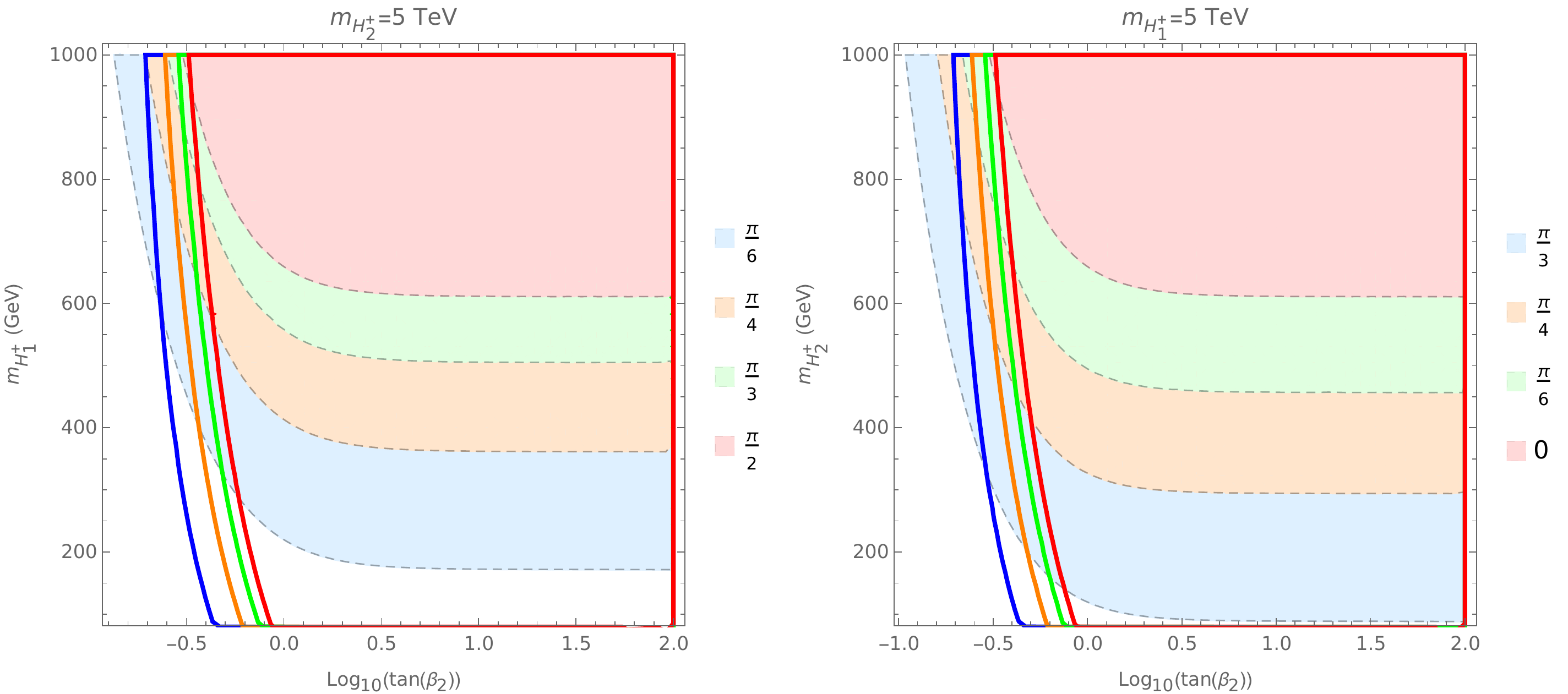}
\caption{
	\small Experimentally allowed regions for the $b \to s \gamma$ branching ratio (colored regions), as well as the boundaries placed by the 
	neutral meson oscillations $\Delta M_{B_s}$ and $\Delta M_{B_d}$, shown by the solid lines. The allowed region for the meson oscillations lies within the 
	boundaries. The color labels denote the value of $\gamma_2$ used in the analysis. The results are shown in the $\tan\beta_2$ {\it vs} the lighter charged Higgs 
	mass plane. \textbf{Left:} $m_{H_2^+} = 5$ TeV, $\tan\beta_1=10$, $\gamma_2 = \{ \pi/6, \pi/4, \pi/3, \pi/2\}$. The 2HDM-II limiting case is $\gamma_2= \pi/2$. 
	\textbf{Right:} $m_{H_1^+} = 5$ TeV, $\tan\beta_1=10$, $\gamma_2 = \{ \pi/3, \pi/4, \pi/6, 0\}$. The 2HDM-II limiting case is $\gamma_2= 0$. 
	Notice the different arrangement of $\gamma_2$ values due to the difference between the trigonometric functions of Eqs.~\eqref{yukla5} and \eqref{yukla6}. 
}
\label{fig:massbounds1}
\end{figure}

\begin{figure}
\centering
\includegraphics[width=0.5\textwidth]{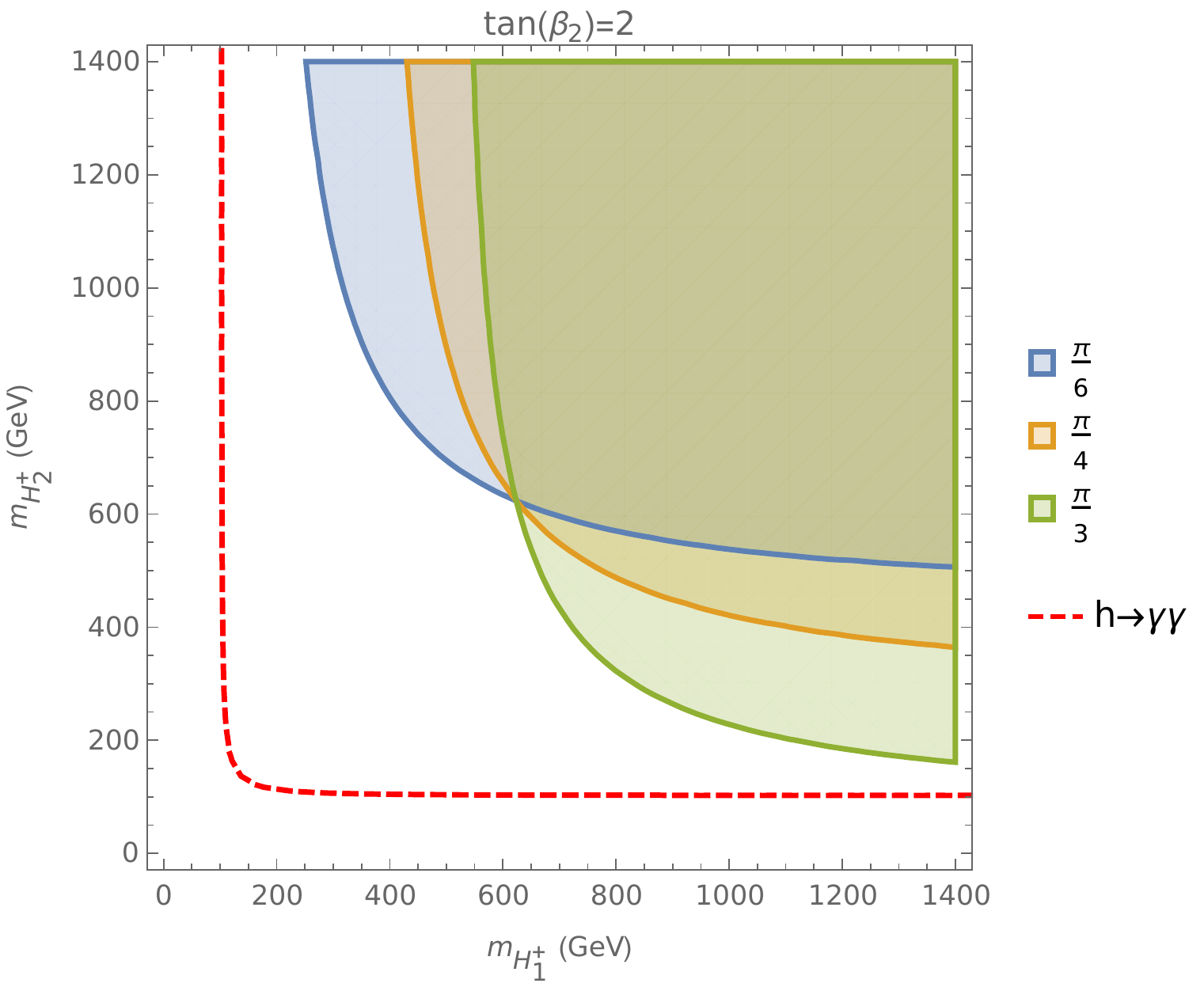}
\caption{\small Experimentally allowed regions at 95\% C.L. from the $b \to s \gamma$ branching ratio (colored regions), where the region of interest is already in agreement with 
$\Delta M_{B_s}$ and $\Delta M_{B_d}$. The color labels denote the values of $\gamma_2$ used in the analysis. The results are shown in the $m_{H_1^+}$-$m_{H_2^+}$ 
plane, and $\tan\beta_1=10$, $\tan\beta_2=2$, $\gamma_2 = \{ \pi/6, \pi/4, \pi/3 \}$. In dashed line we display the $h \to \gamma\gamma$ bounds studied in Sec.~\ref{s:diphoton} where
we set $m_{H_i^+}=m_{H_i}=m_{A_i}$, $i=1,2$. The allowed region at 95\% C.L. from the $h \to \gamma\gamma$ constraint lies above the dashed line.}
\label{fig:massplane}
\end{figure}

Our point is clearly exemplified in Fig.~\ref{fig:massbounds1} where we note that the type-II 2HDM bounds coincide with the more restrictive case of 
this Z3HDM limit ($\gamma_2=\pi/2$ for the bounds on $H_1^\pm$, and $\gamma_2=0$ for $H_2^\pm$).
As we can see, for our benchmark of $\tan\beta_1=10$, the constraints on the charged scalar 
masses are, at worst, comparable to the corresponding bounds in type-II 2HDM for appropriate
values of $\gamma_2$. But the important point is that by changing the values of $\gamma_2$
the bounds can be considerably diluted.
 Even while keeping away from the extremal cases, the bounds can be easily relaxed by 
a factor of 2, by taking $\gamma_2 = \pi/4$, as clearly seen in the plots. 
From Fig.~\ref{fig:massbounds1} we also note that there is an asymmetry in the bounds
on $H_1^\pm$ and $H_2^\pm$ when we are away from the type-II 2HDM limit.
This feature can be attributed to the $\tan\beta_1$ dependency of the charged-Higgs couplings. Moreover, considering the particular nature of the $\tan\beta_2$ dependence of 
both the $b \to s \gamma$ and $\Delta M_{B_{s,d}}$ bounds, we see that for a intermediate range $2 \lesssim \tan\beta_2 \lesssim 30$, the bounds on the charged Higgs masses are
practically independent of $\tan\beta_2$. Thus, by choosing $\tan\beta_2$ in this range, we can lift the assumption of a decoupled charged Higgs, and instead analyze the 
interplay between both contributions to the flavor data, placing the bounds on the 
$m_{H_1^+}$-$m_{H_2^+}$ plane. The results can be seen in Fig.~\ref{fig:massplane}, 
where we show the region compatible with the $b \to s \gamma$ constraints, on the charged Higgs mass plane, while taking $\tan\beta_2 = 2$ as a benchmark. We have checked 
explicitly that the $\Delta M_{B_{s,d}}$ constraints are also satisfied on the region of interest of Fig.~\ref{fig:massplane}, {\it i.e.}, they do not impose additional
restrictions in the $m_{H_1^+}$-$m_{H_2^+}$ plane. The intersection point between all the 
different values of $\gamma_2$ coincides with the type-II 2HDM bound on its charged Higgs mass. Evidently, considerably light charged scalars with masses as low as $\order(200~{\rm GeV})$, can be allowed from flavor data by taking the other charged scalar to be heavier, while still keeping away from extreme values of $\gamma_2$. 

Now that we have established that relatively light charged scalars can successfully pass
through the stringent constraints imposed by the flavor data, it will be interesting if we
can say something about the masses of the neutral nonstandard scalars in relation to those
of the charged scalars. This is where the constraints from the electroweak $\rho$-parameter
become useful. The neutral scalars are expected to have masses such that the impact of NP on
the $\rho$-parameter is minimized. Using the general expressions in Refs.~\cite{Grimus:2007if,Grimus:2008nb} 
we have calculated the NP contribution to the $\rho$-parameter in the
 alignment limit of our model. The relevant expression is particularly clean and
 intuitive in the limit $\gamma_1=\gamma_2=-\alpha_3 =\alpha$ (say) as we display below:
\begin{eqnarray}
\Delta \rho &=& \frac{g^2}{64 \pi^2 m_W^2} \Big\{
F \left( m_{H_1^{+}}^2, m_{A_1}^2 \right)
+ 
F \left( m_{H_2^{+}}^2, m_{A_2}^2 \right)
+
F \left( m_{H_1^{+}}^2, m_{H_1}^2 \right) 
+ 
F \left( m_{H_2^{+}}^2, m_{H_2}^2 \right) \nonumber\\ 
&& \qquad\qquad 
- F \left( m_{A_1}^2, m_{H_1}^2 \right)
-F \left( m_{A_2}^2, m_{H_2}^2 \right)
\Big\} \,,
\label{e:delrho}
\end{eqnarray}
where,
\begin{equation}
F \left( x, y \right) \equiv
\left\{ \begin{array}{ll}
{\displaystyle \frac{x+y}{2} - \frac{xy}{x-y}\, \ln{\frac{x}{y}}}
& \text{for}\ x \neq y,
\\*[3mm]
0 &\text{for}\ x = y.
\end{array} \right.
\end{equation}
One easy way to circumvent the $\rho$-parameter constraint will be to impose $m_{H_1^+}
\approx m_{H_1}\approx m_{A_1} = M_1$ (say) and $m_{H_2^+}
\approx m_{H_2}\approx m_{A_2} = M_2$ (say) as $\Delta\rho$ becomes zero in this limit.
Under this assumption, the scalar spectrum conveniently breaks down into two degenerate 
tiers of nonstandard masses. 
This spectrum of masses and mixings can be easily achieved with a simplified scalar potential of
the following form, which has an enhanced symmetry in its quartic part \cite{Darvishi:2019dbh}:
%
\begin{eqnarray}
\label{e:truncated}
V &=& m_{11}^2(\phi_1^\dagger \phi_1) +  m_{22}^2(\phi_2^\dagger 
\phi_2) +  m_{33}^2 (\phi_3^\dagger \phi_3)  - \left(m_{12}^2(\phi_1^\dagger \phi_2) +  m_{23}^2(\phi_2^\dagger 
\phi_3) +  m_{13}^2 (\phi_1^\dagger \phi_3) + {\rm h.c.} \right)  \\ \nonumber
&& + \lambda(\phi_1^\dagger 
\phi_1 + \phi_2^\dagger \phi_2 + \phi_3^\dagger \phi_3)^2 \,.
\end{eqnarray}
In the above potential there are seven parameters which can be traded in favor of the seven
physical parameters, $(v,\beta_1,\beta_2,m_h,M_1,M_2,\alpha)$. The relevant reparametrizations
are given below:
\begin{subequations}
\label{e:truncinv}
\begin{eqnarray}
s_{12}\equiv \dfrac{2m_{12}^2}{v_1v_2} &=& \frac{2 M_1^2}{v^2} \left[ \frac{c_\alpha^2}{c_{\beta_2}^2} - \tan \beta_2 \left( \frac{c_{2\beta_1} s_{2\alpha}}{s_{2 \beta_1} c_{\beta_2}} + s_{\alpha}^2 \tan \beta_2 \right)\right] \nonumber \\
&& + \frac{2 M_2^2}{v^2} \left[ \frac{s_\alpha^2}{c_{\beta_2}^2} + \tan \beta_2 \left( \frac{c_{2\beta_1} s_{2\alpha}}{s_{2 \beta_1} c_{\beta_2} } - c_{\alpha}^2 \tan \beta_2 \right)\right] \,, \\
s_{13} \equiv \dfrac{2m_{13}^2}{v_1v_3} &=&  \frac{2}{v^2} \left[M_1^2 s_{\alpha}^2 + M_2^2 c_{\alpha}^2 - (M_1^2 - M_2^2)\frac{ c_\alpha s_{\alpha} }{s_{\beta_2} } \tan \beta_1 \right] \,,  \\
s_{23} \equiv \dfrac{2m_{23}^2}{v_2v_3} &=& \frac{2}{v^2} \left[M_1^2 s_{\alpha}^2 + M_2^2 c_{\alpha}^2 + (M_1^2 - M_2^2)\frac{ c_\alpha s_{\alpha} }{s_{\beta_2} \tan \beta_1} \right]\,, \\
\label{e:lambda}
 \lambda &=& \frac{m_h^2}{2 v^2} \,,
\end{eqnarray}
\end{subequations}
where $c_x$ and $s_x$ are shorthands for $\cos x$ and $\sin x$ respectively.

\begin{figure}
\centering
\includegraphics[width=0.6\textwidth]{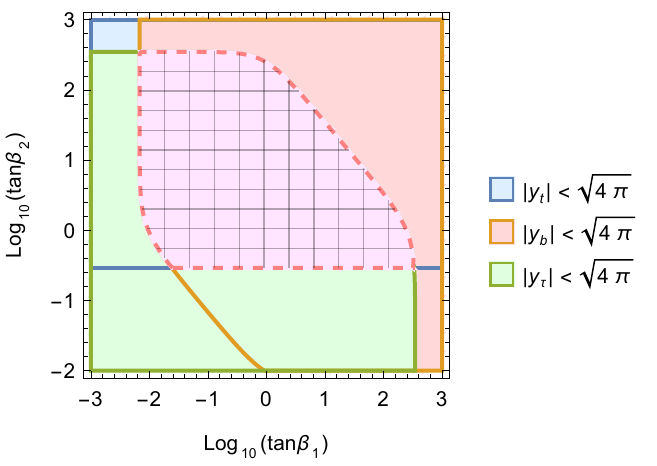}
\caption{
  \small Allowed regions from the perturbativity conditions of the
  Yukawa couplings. The individual color labels
  denote the regions allowed from the top, bottom and $\tau$
  Yukawa couplings and the hatched region represents the combined
  perturbative regime.
}
\label{fig:pert}
\end{figure}

At this point, we wish to remark that the potential of \Eqn{e:truncated} contains only one quartic parameter, $\lambda$.
Thus, both unitarity and stability of the scalar potential can be ensured by requiring $0<\lambda<4\pi$\footnote{For more general analysis of unitarity and boundedness from below conditions for this model, please see Ref.~\cite{Boto:2021qgu}. }.
Looking at \Eqn{e:lambda}, we can easily see that the potential of \Eqn{e:truncated} is manifestly compatible with the unitarity and vacuum stability constraints.
Next, we extract the top, bottom, and $\tau$ Yukawa couplings as 
\begin{eqnarray}\label{eq:yukawa}
  y_t = \frac{\sqrt{2}\, m_t }{v \sin\beta_2}\;,
  \quad y_b = \frac{\sqrt{2}\, m_b }{v \sin\beta_1 \cos\beta_2} \;,
  \quad y_\tau = \frac{\sqrt{2}\, m_\tau }{v \cos\beta_1 \cos\beta_2} \;,
\end{eqnarray}
which follow from our convention that $\phi_3$, $\phi_2$, and $\phi_1$ couple to up-type quarks, down-type quarks, and charged leptons respectively. For the perturbativity of Yukawa couplings, we should have $\lvert y_t \rvert,\lvert y_b\rvert,\lvert y_\tau\rvert < \sqrt{4\pi}$. 
The resulting constraint from perturbativity has been displayed in Fig.~\ref{fig:pert}.
Throughout our paper, we have used values of $\tan\beta_{1,2}$ which are consistent with this perturbative region.

\section{Implications for diphoton decay rate}
\label{s:diphoton}
At this point one might naturally wonder whether such light charged scalars would leave
observable imprints in loop induced Higgs decays such as $h\to \gamma\gamma$.
After the 13 TeV run of the LHC, updated constraints on the Higgs to diphoton signal strength has been reported by
both the ATLAS~\cite{ATLAS:2020pvn} and CMS~\cite{CMS:2020omd} collaborations at  139~$\rm fb^{-1}$ luminosity. It is thus important that we check whether such light charged
scalars can negotiate the bound arising from the measurement of the Higgs to diphoton
signal strength. To do that, we need to calculate the $hH_i^+H_i^-$ couplings which, for
the potential of \Eqn{e:truncated} are given below:
\begin{eqnarray}
g_{h H_i^+ H_i^-} = - \frac{m_h^2}{v} \,, \quad {(i= 1,2)} \label{e:ghh+h-} \,.
\end{eqnarray}
Using this, we can easily write down the expression for the diphoton signal strength
as follows:
\begin{eqnarray}
	\mu_{\gamma\gamma} = \frac{|F_W(\tau_W) + \frac{4}{3} F_t(\tau_t) +\sum_{i=1}^{2} \kappa_i F_{i+}(\tau_{i+}) |^2}{|F_W(\tau_W) + \frac{4}{3} F_t(\tau_t) |^2} \,,
\end{eqnarray}
	where, 
	$ \kappa_i = -{m_h^2}/{2 m_{H_i^+}^2}$,  $\tau_x =\left( {2 m_x}/{m_h}\right)^2$,  $\left(x = W, t, H_i^+\right)$ and the loop functions are given by~\cite{Gunion:1989we},
 \begin{subequations}
	\begin{eqnarray}
	F_W(x) &=& 2 + 3 x + 3 x (2 - x)f(x) \,,  \\
	F_t(x) &=& - 2x\left[1 + (1 - x) f(x)\right] \,,  \\
	F_{i+} (x)&=& - x \left[1 - x f(x)\right] \,.
	\end{eqnarray}
\end{subequations}
with, $f(x) = \left[\sin^{-1}\left(\sqrt{1/x}\right)\right]^2$ for $x > 1$.
It is interesting to note that in the limiting potential of \Eqn{e:truncated}, the charged Higgs couplings to the SM-like Higgs in \Eqn{e:ghh+h-} are completely independent of any mixing angles and fixed to a constant value. Therefore, the charged Higgs contribution to the loop-induced Higgs to diphoton channel will always be suppressed by the charged Higgs masses when the charged Higgses are much larger than the SM-like Higgs.
We display our results in Fig.~\ref{fig:massplane} in the $(m_{H_1^+}$-$m_{H_2^+} )$ mass plane, 
where we see that the current Higgs data  mainly discards the parameter space where both or any one of the charged Higgs masses are below  $\order(200~{\rm GeV})$. In Fig.~\ref{fig:massplane}, the region below the red dashed line is excluded by the current data at 95\% C.L~\cite{ATLAS:2020pvn}.
%

\section{Direct search constraints}
\label{s:direct}
The presence of two charged and four additional neutral
Higgs bosons places this model under the scrutiny
of direct searches for nonstandard Higgs bosons at the LHC.
In the parameter region favored by the flavor physics constraints
($m_{H_{1,2}^+} > m_{\rm top}$), the  dominant production mode of a charged
Higgs boson at the LHC is in association with $tb$-quark pairs.
Both ATLAS and CMS collaborations have provided model-independent
upper bounds on the production cross-section
times branching ratio for this mode with the charged Higgs boson
decaying to $tb$~\cite{Aad:2021xzu,Sirunyan:2019arl} and \
$\tau \nu$~\cite{Aaboud:2018gjj,Sirunyan:2019hkq} final states\footnote{
The vanishing $H_1^\pm W^\mp h$ coupling in the alignment limit leads to the
absence of $W^\pm h$ final state in $H_1^\pm$ decay.
Furthermore, the decay to $W^\pm A_1$ final state is kinematically disfavored
as a result of the assumed degnearcy between $H_1^\pm$ and $A_1$.}.
On the other hand, the search for heavy scalar and pseudoscalar
resonances yields the most stringent constraints in the $\tau^+\tau^-$
final state. In this case model-independent
bounds are available for production via gluon-gluon-fusion and in
association with a b-quark pair~\cite{Aad:2020zxo,Sirunyan:2018zut}. In the following, we discuss the
impact of the various bounds mentioned above
on the parameter space of Z3HDM.
\begin{figure}[ht]
	\centering
	\begin{subfigure}[b]{0.32\linewidth}
		\centering\includegraphics[width=\textwidth]{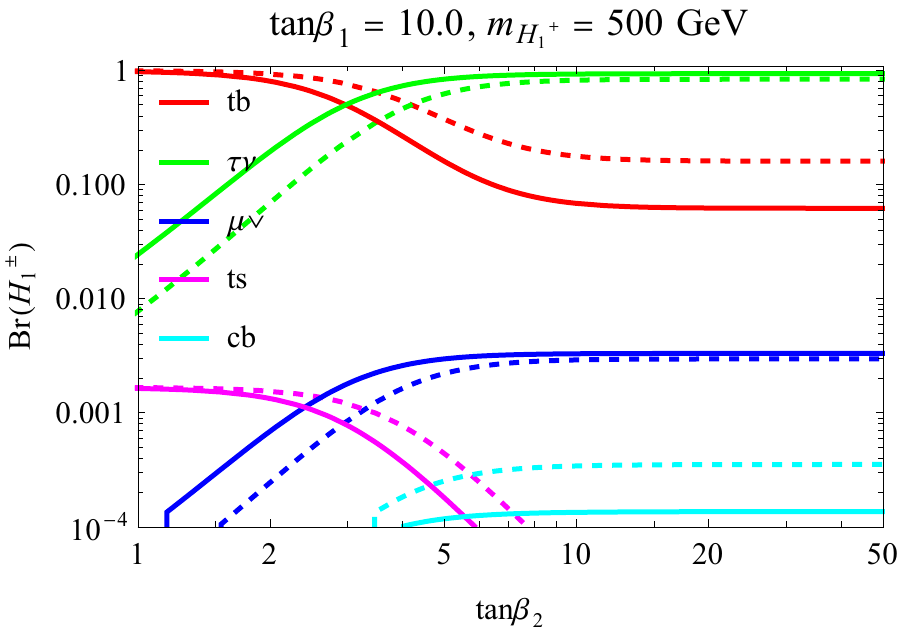}
		\caption{}
		\label{fig:h1pmbr}
	\end{subfigure}
	\begin{subfigure}[b]{0.32\linewidth}
		\centering\includegraphics[width=\textwidth]{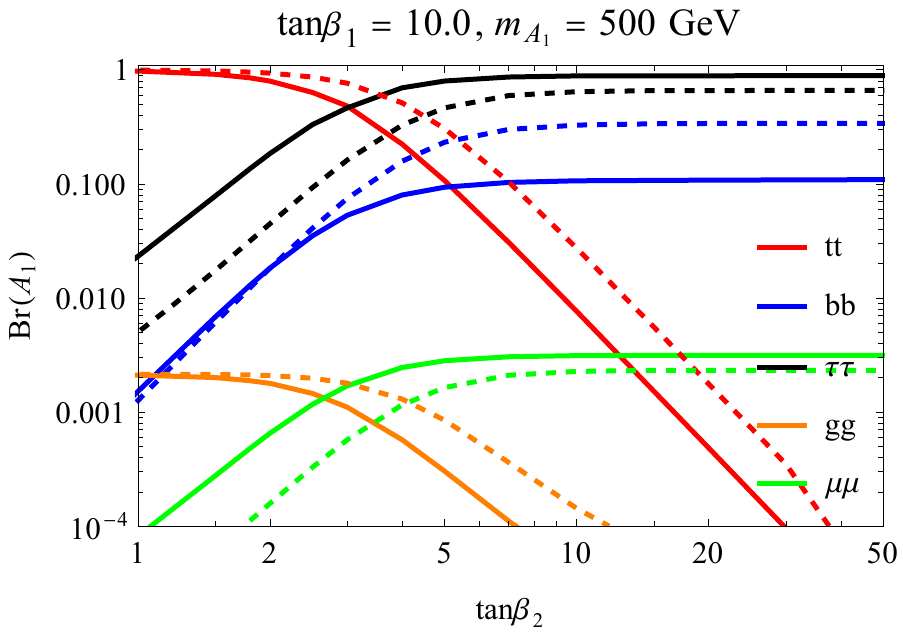}
		\caption{}
		\label{fig:a1br}
	\end{subfigure}
	\begin{subfigure}[b]{0.32\linewidth}
		\centering\includegraphics[width=\textwidth]{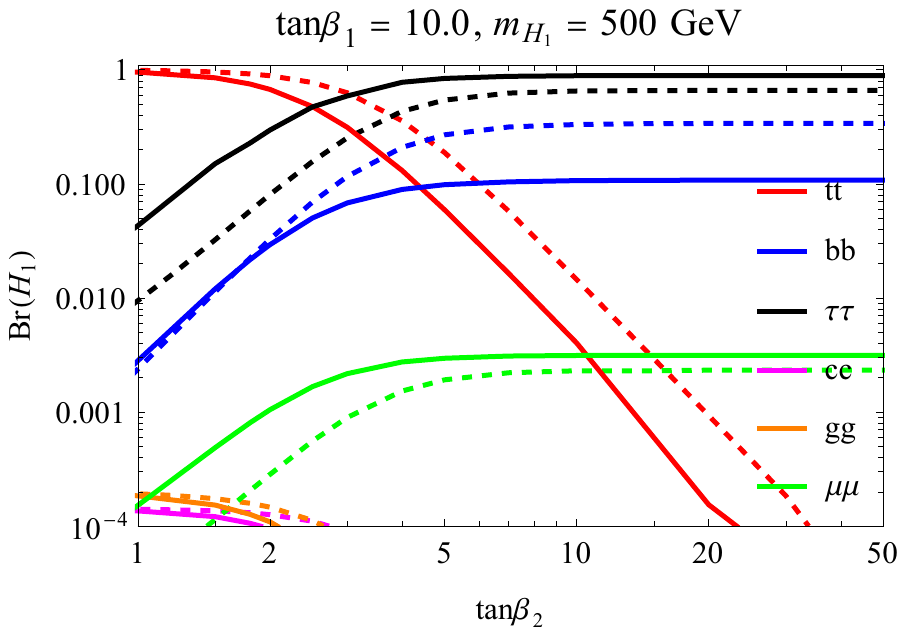}
		\caption{}
		\label{fig:h1br}
	\end{subfigure}
	\caption{\small The figure shows the branching fractions of $H_1^\pm$, $A_1$ and $H_1$ 
		with $\tan\beta_2$ for fixed masses of
		$m_{A_1} = m_{H_1} = m_{H_1^+} = 500$ GeV.
		Here we choose, $\gamma_1=\gamma_2=-\alpha_3 = \frac{\pi}{6}$ (solid lines)
		and $\frac{\pi}{4}$ (dashed Lines) and $m_{A_2} = m_{H_2} = m_{H_2^+} = 5$ TeV.
                { This hierarchy of branching fractions can serve as distinguishing feature
                of Z3HDM from other canonical 2HDMs~\cite{Aoki:2009ha}.}}
	\label{fig:br}
\end{figure}

To elucidate the relevance of different direct search constraints on the parameter
region of our interest, we plot in Fig.~\ref{fig:br} the branching ratios of
$H_1^\pm$, $A_1$ and $H_1$ as a function of $\tan\beta_2$.
Keeping in mind the precision constraints from electroweak $\rho$-parameter,
we choose to work in the limit $m_{H_i^+}=m_{H_i}=m_{A_i}$, $i=1,2$
and $\gamma_1=\gamma_2=-\alpha_3$. Furthermore, we consider the case of 
one decoupled charged Higgs boson for simplicity,
which in this case we take to be $H_2^\pm$ \footnote{Similar 
bounds may be imposed on $H_2^\pm$ in the case of a decoupled $H_1^\pm$.}.
The branching ratios are calculated for a fixed $\tan\beta_1=10$ and
two different values of $\sin\gamma_2$ shown as solid ($\gamma_2 =\frac{\pi}{6}$) and
dashed ($\gamma_2=\frac{\pi}{4}$) lines. For $\tan\beta_2 \lesssim 1$
the leading decay mode of $H_1^\pm$ is $H_1^\pm \to tb$,
because of the dominance of the first term 
in Eq.~\eqref{yukla5}, which is proportional to $\cot\beta_2$. 
As $\tan\beta_2$ increases, the terms proportional to
$\tan\beta_2$ in Eqs.~\eqref{yukla4} and \eqref{yukla5} promptly takes over.
In the $\tan\beta_2 >1$ region favored by the flavor physics data,
$H_1^\pm \to \tau\nu$ becomes the dominant decay mode.
A somewhat similar pattern is observed in the branching ratio of $A_1$ and $H_1$,
with $A_1/H_1\to\tau^+\tau^-$ being the dominant decay mode in the $\tan\beta_2 >1$ region.

\begin{figure}[ht]
  \centering
\begin{subfigure}[b]{0.32\linewidth}
\centering\includegraphics[width=\textwidth]{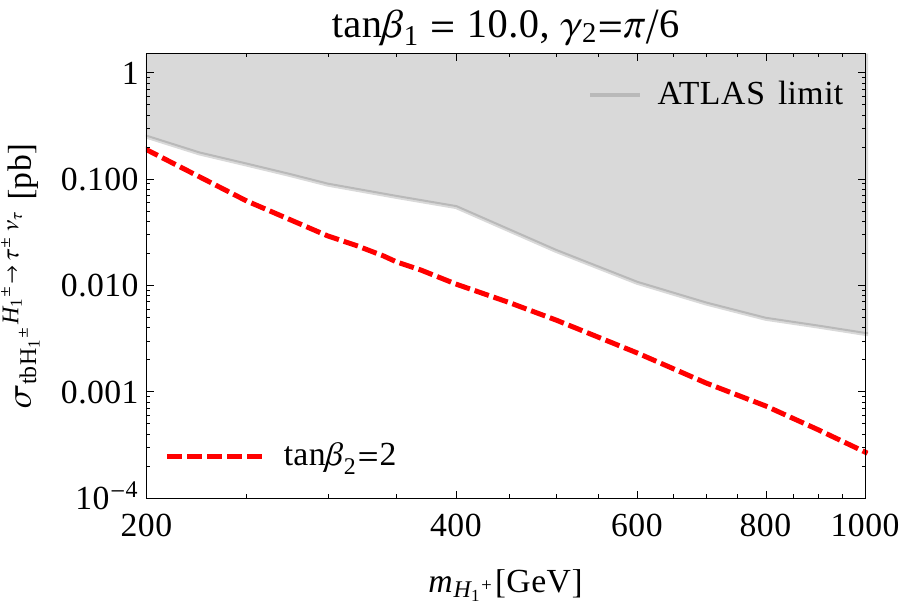}
        \caption{}
        \label{fig:h1pmsigmabr}
\end{subfigure}
\begin{subfigure}[b]{0.32\linewidth}
\centering\includegraphics[width=\textwidth]{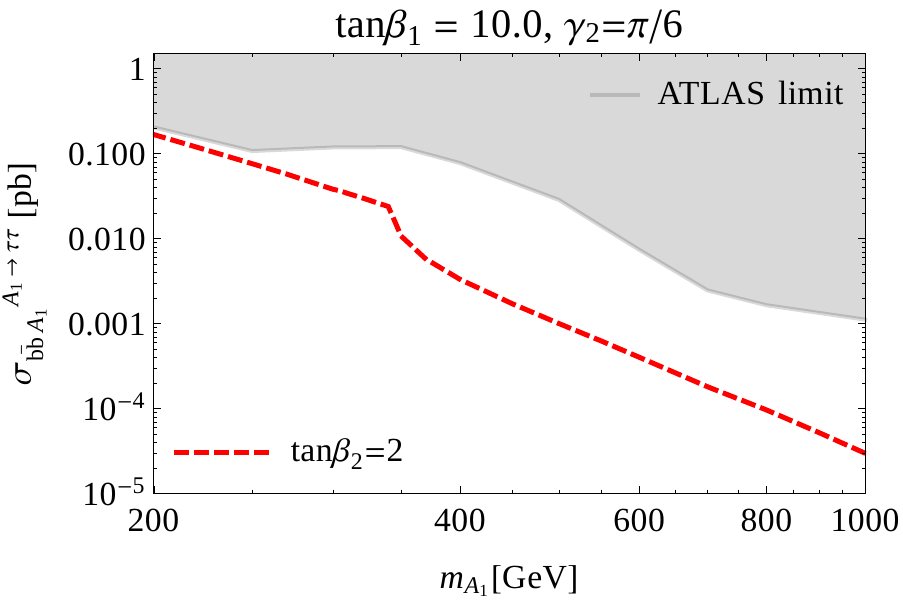}
        \caption{}
        \label{fig:a1sigmabr}
\end{subfigure}
\begin{subfigure}[b]{0.32\linewidth}
\centering\includegraphics[width=\textwidth]{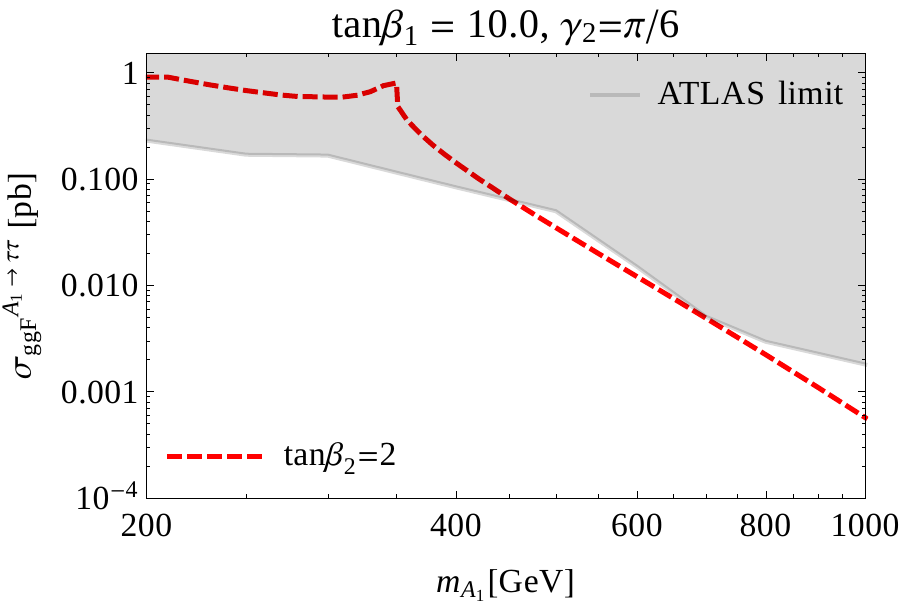}
        \caption{}
        \label{fig:h1sigmabr}
\end{subfigure}
\caption{\small The figure shows the cross-section times branching ratio of
  $H_1^\pm$ and $A_1$ at the 13 TeV LHC as a function of their masses.
  The plots correspond to the parameter choice $\tan\beta_1=10, \tan\beta_2=2,
  \gamma_1=\gamma_2=-\alpha_3 = \frac{\pi}{6}$ and $m_{A_2} = m_{H_2} = m_{H_2^+} = 5$ TeV.
  $\sigma^x_y$ denotes the production cross-section times branching ratio
  for the production mode $x$ and decay mode $y$ of the corresponding Higgs boson.
  The parameter space excluded by the latest bound from ATLAS is shown
  as a gray-shaded region.}
\label{fig:sigmabr}
\end{figure}

The implications of various direct search constraints on the
parameter space of Z3HDM becomes even clearer by looking at
Fig.~\ref{fig:sigmabr} where we show the production cross-section
times branching ratio of the Higgs bosons $H_1^\pm, A_1$ and $H_1$
at the 13 TeV LHC as a function of their masses.
For this calculation, we implemented our model in
\texttt{FeynRules}~\cite{Christensen:2008py,Alloul:2013bka}
to generate files in the \texttt{UFO} format~\cite{Degrande:2011ua}.
These files are then used by {\tt MadGraph5\_aMC@NLO}~\cite{Alwall:2014hca} to compute
the signal cross-section at the LHC. The gray-shaded region denotes
the parameter space excluded by the corresponding bound from ATLAS.
We consider a relatively
small value of $\tan\beta_2$,  namely, $\tan\beta_2 = 2$, to comply
with the stringent bounds from the LHC. The value of $\tan\beta_1$ is kept
fixed at $\tan\beta_1=10$.
As can be seen from Fig.~\ref{fig:h1pmsigmabr}, the value of
$\sigma^{{H_1^\pm \to \tau\nu}}_{tbH_1^\pm}$ remains comfortably within the upper limit
set by the direct search for charged Higgs boson mass within the region of interest.
Thus, even the relatively low $m_{H_1^+}$
region allowed by the flavor physics constraints remains effectively safe
from the direct LHC constraints on charged Higgs bosons. However, the bounds from 
scalar and pseudoscalar resonance searches in the ditau channel can be more
constraining in this case, especially when taken together with the indirect
bounds from the precision measurements of the electroweak
$\rho$-parameter and diphoton decay rate of the 125 GeV Higgs boson as
discussed in Sec.~\ref{s:diphoton}.
It can be clearly seen from Fig~\ref{fig:a1sigmabr} that the 
associated production of the pseudoscalar $A_1$
with b-quark pairs does not impose any significant restrictions
on the relevant parameter space. However, the production cross-section via
gluon-gluon-fusion (ggF) process, as shown in Fig.~\ref{fig:h1sigmabr},
can be significantly larger in this case,
allowing only $m_{A_1}\gtrsim 450$~GeV.
The similar LHC bounds on $H_1$ comes out to be weaker than those on $A_1$ in most of
the parameter space for this scenario.

Before we conclude, it should be noted that this seemingly strong constraint from the 
direct searches should be interpreted with some
caution. The strong constraints on the neutral bosons are essentially due to relatively
high BRs into the $\tau \tau$ channel, which, in turn, may be attributed to our choice of
$\tan\beta_1\gg 1$. But one should also remember that we have been working in a simplified
limit of Z3HDM where the nonstandard scalars come with two tiers of degenerate masses, which
we motivated from the $\rho$-parameter constraints. However, we can lift the degeneracy and
allow one of the neutral scalars to have a different mass while still keeping the NP
contributions to the $\rho$-parameter under control. In this way, it will be possible
to open up channels like $A_1\to H_1 Z$ which will reduce BR$(A_1\to \tau\tau)$. Moreover,
there can be additional decay modes in the scalar sector too. As an example, if there is
a dark singlet coupling to the other scalars, then decay modes such as $H_1\to SS$ where
$S$ is the dark singlet, can open up. Keeping all these possibilities in mind, we can say
that the bounds from the direct searches in our simplified analysis can be considerably
relaxed.

\section{Summary}
\label{s:summary}
To summarize, we have analyzed a 3HDM with $Z_3$-symmetry featuring NFC,
where the $Z_3$-symmetry ensures a democratic Yukawa structure requiring each type of
SM fermion to be coupled to a particular Higgs scalar doublet, thus eliminating
FCNCs at tree level. We have discussed the characteristics of the scalar
and Yukawa sectors in detail, focusing on the alignment limit where the lightest
CP-even Higgs boson of the model possesses SM-like tree-level couplings and hence
can serve as a candidate for the 125 GeV scalar observed at the LHC. This alignment
limit can be characterized by a set of simple analytic conditions closely resembling that
of a 2HDM. The alignment limit is also phenomenologically well-motivated
in view of the increasingly precise measurements of the signal
strengths of the 125 GeV SM-like Higgs boson at the LHC.

The presence of two pairs of charged ($H_{1,2}^\pm$) and 
additional two neutral CP-odd ($A_{1,2}$) and two neutral CP-even ($H_{1,2}$) Higgs
bosons in the model gives rise to distinctive signatures
in various experiments looking for direct or indirect signals of BSM physics.
From the phenomenological point of view, we have put an emphasis on analyzing the
effect of the flavor physics constraints on the parameter space of Z3HDM.
The leading BSM contribution to flavor observables like BR$(b \to s\gamma)$ and
$\Delta M_{B_{s,d}}$ comes from the loops containing the charged Higgs bosons $H_1^\pm$
and $H_2^\pm$. The Yukawa coupling structure  of the charged Higgs
bosons in this model bear close likeness to those of the Type-II 2HDM.
However, the key difference from Type-II 2HDM is 
that the fermionic couplings of the charged Higgses feature an
additional suppression effect which is essentially non-decoupling
in nature. Thus, even in the limit of an effective Type-II 2HDM, with one of the charged Higgs
bosons taken to be decoupled from the spectrum, the couplings of the other charged Higgs retains
the suppression factor. This produces a significant relaxation of the bounds coming from flavor
observables in this model compared to Type-II 2HDM. It is observed that charged Higgs masses as low
as 200 GeV are allowed by the flavor data in Z3HDM, whereas in the case of Type-II 2HDM
the lower bound on charged Higgs mass from the same flavor physics constraints
stands at $\order(600~{\rm GeV})$.
We have discussed the combined contribution
of $H_1^\pm$ and $H_2^\pm$ to flavor observables when both of them are taken to be light.
We have also taken into account the precision constraints from
EW $\rho$-parameter which can be easily satisfied in a simple set up with two sets of
degenerate masses for the nonstandard scalars, $m_{H_i^+}=m_{H_i}=m_{A_i}$, $i=1,2$.
We show that in this limiting case, the couplings of the charged Higgs bosons to
the 125 GeV SM-like Higgs assumes a constant value. Therefore, the contribution
to $h\to\gamma\gamma$ decay from charged Higgs loop, being suppressed by
a factor of ${m_h^2}/{m_{H_i^+}^2}, i=1,2$, does not produce any additional constraint
on the relevant parameter space.

Finally, we have analyzed the constraints coming from the direct searches of the nonstandard
Higgs bosons at the LHC. We show that the bounds from direct charged Higgs boson searches can
be satisfied relatively easily in most of the parameter region satisfied by the flavor data.
However, the constraints on neutral CP-even and CP-odd Higgs boson masses coming
from ditau resonance searches put somewhat stringent bounds on our parameter space.
We also note that the strong constraints
from the ditau resonance searches are a consequence of the particular parameter choice
we make to satisfy the bounds from EW $\rho$-parameter.
Some alternative choice of parameters to satisfy the
$\rho$-parameter constraint may lead to a significant reduction in the branching ratios
of the neutral CP-even and CP-odd scalars to $\tau\tau$ final states.
This, in turn, may lead to a dilution of the LHC bounds by a considerable amount.

\section*{Acknowledgements}
DD thanks the Science and Engineering Research Board, India for financial support
through grant no. SRG/2020/000006.
The work of IS was supported by World Premier International Research Center Initiative (WPI), MEXT, Japan.
The work of MC is supported by the project AstroCeNT:
Particle Astrophysics Science and Technology Centre,  carried out within
the International Research Agendas programme of
the Foundation for Polish Science financed by the
European Union under the European Regional Development Fund.
 The work of ML is funded by
Funda\c{c}\~{a}o para a Ci\^{e}ncia e Tecnologia-FCT through Grant
No.PD/BD/150488/2019, in the framework of the Doctoral Programme
IDPASC-PT, and in part by the projects CFTP-FCT Unit 777 
(UIDB/00777/2020 and UIDP/00777/2020), and CERN/FIS-PAR/0008/2019.  

\appendix

\section{Appendix}
\label{app:ScalarSector}
In this Appendix we give a detailed description of the scalar sector.
It is to be noted that the scalar potential of \Eqn{e:potential}
contains 18 parameters including the three soft-symmetry breaking terms $m_{12}^2$, $m_{13}^2$ and $m_{23}^2$. 
Among all, the bilinear parameters
$m_{11}^2$, $m_{22}^2$ and $m_{33}^2$ can be traded for the three
VEVs, $v_1$, $v_2$ and $v_3$ or equivalently $v$, $\tan\beta_1$
and $\tan\beta_2$.The remaining twelve quartic couplings will correspond to
the seven physical masses (three CP-even scalars,
two CP-odd scalars and two pairs of charged scalars) and five
mixing angles (three in the CP-even sector, one in the CP-odd
sector and one in the charged scalar sector). Below, we demonstrate this
relations by examining the potential of \Eqn{e:potential}
in more detail.

The minimization conditions used to replace the
bilinear parameters in terms of the VEVs are given below:
\begin{subequations}
    \begin{eqnarray}
    m_{11}^2 &=&   -\lambda_1 v_1^2 -\frac{1}{2}\left\{
    (\lambda_4+\lambda_7)v_2^2 +(\lambda_5+\lambda_8)v_3^2
    +2\lambda_{10}v_2 v_3\right\}\nonumber\\ && -\frac{v_2v_3}{2v_1}
    \left(\lambda_{11}v_2 +\lambda_{12}v_3 \right) + m_{12}^2 \frac{v_2}{v_1} + m_{13}^2 \frac{v_3}{v_1} \,, 
     \\
    m_{22}^2 &=& -\lambda_2 v_2^2 -\frac{1}{2}\left\{
    (\lambda_4+\lambda_7)v_1^2 +(\lambda_6+\lambda_9)v_3^2
    +2\lambda_{11}v_1 v_3\right\} \nonumber\\ && -\frac{v_1v_3}{2v_2}
    \left(\lambda_{10}v_1 +\lambda_{12}v_3 \right) + m_{12}^2 \frac{v_1}{v_2} + m_{23}^2 \frac{v_3}{v_2}\,, 
    \\
    m_{33}^2 &=& -\lambda_3 v_3^2 -\frac{1}{2}\left\{
    (\lambda_5+\lambda_8)v_1^2 +(\lambda_6+\lambda_9)v_2^2
    +2\lambda_{12}v_1 v_2\right\} \nonumber\\ && -\frac{v_1v_2}{2v_3}
    \left(\lambda_{10}v_1 +\lambda_{11}v_2 \right) + m_{13}^2 \frac{v_1}{v_3} + m_{23}^2 \frac{v_2}{v_3}\,. 
    \end{eqnarray}
    \label{e:bilinears}
\end{subequations}
Now let us demonstrate the diagonalization of the mass matrices in different sectors 
following the same prescription as in \cite{Das:2019yad}
but in the presence of the soft terms.


\subsection{CP-odd scalar sector} 
The mass term for the  pseudoscalar sector can be extracted from the scalar potential as,
		\begin{eqnarray}\label{e:cpoddmassmat}
		V_{PS}^{\rm mass} = \begin{pmatrix}
		z_1 & z_2 & z_3
		\end{pmatrix} \, \frac{{\cal M}_P^2}{2} \, \begin{pmatrix}
		z_1\\  z_2\\ z_3\\
		\end{pmatrix} \,,
		\end{eqnarray}
 where ${\cal M}_{P}^2$ is the $3\times3$ mass matrix that can be block diagonalized as follows:
%
	\begin{subequations}
		\begin{eqnarray}\label{e:mp2x2}
		({\cal B}_{P})^2 \equiv {\cal O}_{\beta} \cdot {\cal M}_{P}^2 \cdot {\cal O}_{\beta}^{T} &=& \begin{pmatrix}
		0 & 0 & 0 \\
		0 & {({\cal B}_P^2)}_{22}  & {({\cal B}_P^2)}_{23} \\
		0 & {({\cal B}_P^2)}_{23} & {({\cal B}_P^2)}_{33} \\
		\end{pmatrix} \,.
		\end{eqnarray}
The elements of ${{\cal B}_P^2}$ are given by,
		\begin{eqnarray}
		{({\cal B}_P^2)}_{22} &=& -\frac{v_3}{2v_1 v_2\left(v_1^2+v_2^2\right)}  \left[ \lambda_{10} v_1\left(v_1^2+2 v_2^2\right)^2+\lambda_{11} v_2\left(2 v_1^2+v_2^2\right)^2+\lambda_{12} v_3 \left(v_1^2-v_2^2\right)^2 +  m_{23}^2 v_1^3 + m_{13}^2 v_2^3  \right] \nonumber \\
		&& +  m_{12}^2\frac{(v_1^2 + v_2^2)}{v_1 v_2} \,, 
		\\
		{({\cal B}_P^2)}_{23}  &=&\frac{v }{2 \left(v_1^2+v_2^2\right)} \left[ -\lambda_{10} v_1 \left(v_1^2 + 2 v_2^2\right)+ \lambda_{11}v_2 \left(2 v_1^2 +v_2^2\right)+ 2\lambda_{12} v_3 \left(v_1^2-v_2^2\right)- m_{23}^2 v_1 +  m_{13}^2 v_2 \right] \,,\\ 
		{({\cal B}_P^2)}_{33} &=&  -\frac{v^2}{2 v_3\left(v_1^2+v_2^2\right)}  \left[\lambda_{10}v_1^2 v_2+ \lambda_{11} v_1 v_2^2+ 4\lambda_{12} v_1 v_2v_3 -  m_{12}^2 v_1 -  m_{23}^2 v_2\right] \,.
		\end{eqnarray}\label{e:BP2}
	\end{subequations}

The matrix ${{\cal B}_P^2}$ can be fully diagonalized by an orthogonal transformation
	\begin{eqnarray}
	{\cal O}_{\gamma_1} \cdot ({\cal B}_{P})^2 \cdot {\cal O}_{\gamma_1}^T &=& 
	{\rm diag}  (0,~ m^2_{A_1},~ m^2_{A_2})
	\,,\label{e:PSrot} 
	\end{eqnarray}
where, ${\cal O}_{\gamma_1}$ is given in Eq.~(\ref{eq:gammaRots}), 
which entails the following relations
\begin{subequations}\label{e:mAtoBP2}
\begin{eqnarray}
m^2_{A_1} \cos^2 \gamma_1 +  m^2_{A_2} \sin^2 \gamma_1 &=& {({\cal B}_P^2)}_{22}   \,, \\
\cos \gamma_1 \sin\gamma_1 (m^2_{A_2} - m^2_{A_1})  &=&	{({\cal B}_P^2)}_{23} \,, \\
m^2_{A_1} \sin^2 \gamma_1 +  m^2_{A_2} \cos^2 \gamma_1 &=& {({\cal B}_P^2)}_{33}  \,.
\end{eqnarray}
\end{subequations}
Using Eq.~(\ref{e:BP2}), Eq.~(\ref{e:mAtoBP2}) can be inverted to solve for $\lambda_{10}\,, \lambda_{11}$ and $\lambda_{12}$ as 
\begin{subequations}\label{e:lam12}
	\begin{eqnarray}
	 \lambda_{10} &=&
	 \frac{2m_{A_1}^2}{9 v^2} \left[\frac{s_{2\gamma_1}}{c_{\beta_1} c_{\beta_2}}-\frac{2s_{\beta_1}  c^2_{\gamma_1}}{s_{\beta_2} c_{\beta_2}}+\frac{s_{3\beta_1} s_{\gamma_1} c_{\gamma_1}}{s_{\beta_1} c_{\beta_1}c_{\beta_2}}+\tan\beta_2 s^2_{\gamma_1} \left\{\frac{\tan\beta_1}{c_{\beta_1}}-2c_{\beta_1} \cot\beta_1\right\}\right] \nonumber \\
	 &&-\frac{m_{A_2}^2}{9 v^2} \left[(2 c_{2\beta_1}+3)\frac{s_{2\gamma_1}}{ c_{\beta_1} c_{\beta_2} }+4\frac{s_{\beta_1}  s^2_{\gamma_1}}{s_{\beta_2} c_{\beta_2}}-2 \tan\beta_2 c^2_{\gamma_1} \left\{\frac{\tan\beta_1}{c_{\beta_1}}-2 c_{\beta_1}\cot \beta_1\right\}\right] \nonumber \\
	 && + \frac{4}{9 v^2}\frac{m_{12}^2}{ s_{\beta_2} c_{\beta_1} c_{\beta_2} } +  \frac{4}{9 v^2}\frac{m_{13}^2}{s_{\beta_2} c_{\beta_1} c_{\beta_2}^2 } -  \frac{2}{9 v^2}\frac{m_{23}^2}{ c_{\beta_1}^2 c_{\beta_2}^2 } \,. \\ 
	\lambda_{11} &=& \frac{m_{A_1}^2}{9 v^2} \left[ - \frac{4 c_{\beta_1} c_{\gamma_1}^2}{s_{\beta_2}c_{\beta_2}} + \frac{(-3 + 2 c_{2\beta_1})}{s_{\beta_1}c_{\beta_2}} s_{2\gamma_1} + 2 \left(\cot^4 \beta_1 + \cot^2 \beta_1-2\right)s_{\beta_1}s_{\gamma_1}^2 \tan \beta_1 \tan \beta_2 \right] \nonumber \\
	&&+\frac{m_{A_2}^2}{9 v^2}\left[ - \frac{4 c_{\beta_1} s_{\gamma_1}^2}{s_{\beta_2}c_{\beta_2}} + \frac{(5 + \cot^2\beta_1)}{c_{\beta_2}} s_{2\gamma_1}s_{\beta_1} + 2 \left(\cot^4\beta_1 + \cot^2 \beta_1-2\right)s_{\beta_1}c_{\gamma_1}^2 \tan \beta_1 \tan \beta_2 \right] \nonumber \\
	&& + \frac{4}{9 v^2}\frac{m_{12}^2}{ s_{\beta_1} s_{\beta_2} c_{\beta_2} } - \frac{2}{9 v^2}\frac{m_{13}^2}{s_{\beta_1}^2 c_{\beta_2}^2 } + \frac{4}{9 v^2}\frac{m_{23}^2}{ s_{\beta_1} c_{\beta_1} c_{\beta_2}^2 } \,. \\ 
	\lambda_{12}  &=& \frac{m_{A_1}^2}{36 v^2} \left[\frac{4 s_{2\beta_1}  c^2_{\gamma_1}}{s ^2_{\beta_2}}-\frac{4 c_{2\beta_1}  s_{2\gamma_1}}{s_{\beta_2}}+(c_{4\beta_1}-17) 
	\frac{s^2_{\gamma_1}}{s_{\beta_1} c_{\beta_1}}\right] \nonumber \\
	&&+\frac{m_{A_2}^2}{36 v^2} \left[ \frac{4s_{2\beta_1} s^2_{\gamma_1}}{ s^2_{\beta_2}}+\frac{4c_{2\beta_1} s_{2\gamma_1}}{s_{\beta_2}} +(c_{4\beta_1}-17) \frac{ c^2_{\gamma_1}}{s_{\beta_1} c_{\beta_1}}\right] \nonumber \\
	&& -  \frac{2 m_{12}^2}{9 v^2  s_{\beta_2}^2 } +  \frac{4}{9 v^2}\frac{m_{13}^2}{ s_{\beta_1} s_{\beta_2} c_{\beta_2} } + \frac{4}{9 v^2}\frac{m_{23}^2}{ c_{\beta_1} s_{\beta_2} c_{\beta_2} } \,.
	\end{eqnarray}
\end{subequations}
%


\subsection{Charged scalar sector} 
Similar to the pseudoscalar case, the $3\times3$ charged sector mass matrix ${\cal M}_{C}^2$ can also be block diagonalized as:
\begin{subequations}\label{e:BC2}
	\begin{eqnarray}
	({\cal B}_{C})^2 \equiv {\cal O}_{\beta} \cdot {\cal M}_{C}^2 \cdot {\cal O}_{\beta}^{T} &=& \begin{pmatrix}
	0 & 0 & 0 \\
	0 & {({\cal B}_C^2)}_{22}  & {({\cal B}_C^2)}_{23} \\
	0 & {({\cal B}_C^2)}_{23} & {({\cal B}_C^2)}_{33} \\
	\end{pmatrix} \,.
	\end{eqnarray}
	where,
	\begin{eqnarray}
	{({\cal B}_C^2)}_{22} &=& -\frac{1}{2(v_1^2+v_2^2)} \bigg[\lambda_{10}\frac{v_3}{v_2} \left(\left(v_1^2+v_2^2\right)^2+v_2^4\right) + \lambda_{11}\frac{v_3}{v_1} \left(\left(v_1^2+v_2^2\right)^2+v_1^4\right) + \lambda_{12}\frac{v_3^2}{v_1 v_2} \left(v_1^4+v_2^4\right)\nonumber \\
	&& + \lambda_7 \left(v_1^2+v_2^2\right)^2+ \lambda_8 v_2^2 v_3^2+ \lambda_9 v_1^2 v_3^2 
	 -  m_{12}^2\frac{(v_1^2 + v_2^2)^2}{v_1 v_2} - m_{13}^2\frac{ v_2^2 v_3 }{v_1} - m_{23}^2\frac{ v_1^2 v_3 }{v_2}  \bigg]\,. \\
	{({\cal B}_C^2)}_{23}  &=& \frac{v}{2(v_1^2+v_2^2)} \left[- v_1 v_2^2 \lambda_{10} + \lambda_{11} v_1^2 v_2 + \lambda_{12} v_3(v_1^2 - v_2^2) \right. \nonumber\\
	&& \qquad \left. -\lambda_8 v_1 v_2 v_3 + \lambda_9 v_1 v_2 v_3  - m_{23}^2 v_1 + m_{13}^2 v_2 \right] \,, \\
    {({\cal B}_C^2)}_{33} &=& -\frac{v^2}{2(v_1^2+v_2^2)} \left[\frac{v_1^2 v_2}{v_3}\lambda_{10} + \lambda_{11} \frac{v_1 v_2^2}{v_3} + 2 v_1 v_2 \lambda_{12} + \lambda_8 v_1^2 + \lambda_9 v_2^2  - m_{13}^2 \frac{v_1}{v_3}  - m_{23}^2 \frac{v_2}{v_3} \right] \,.
	\end{eqnarray}
\end{subequations}

Further, the charged scalar mass matrix can be completely diagonalized  with the use of the rotation matrix ${\cal O}_{\gamma_2}$ (given in 
Eq.~(\ref{eq:gammaRots})) as 
\begin{equation}
{\cal O}_{\gamma_2} \cdot ({\cal B}_{C})^2 \cdot {\cal O}_{\gamma_2}^T = {\rm diag} (0,~ m^2_{H^+_1},~ m^2_{H^+_2})
\end{equation}
Thus, we will have the following relations:
\begin{subequations}\label{e:mctoapbpcp}
\begin{eqnarray}
m^2_{H_1^+} \cos^2 \gamma_2 +  m^2_{H_2^+} \sin^2 \gamma_2 &=& {({\cal B}_C^2)}_{22}\,, \\
\cos \gamma_2 \sin\gamma_2 (m^2_{H_2^+} - m^2_{H_1^+}) &=&  {({\cal B}_C^2)}_{23}\,, \\
m^2_{H_1^+} \sin^2 \gamma_2 +  m^2_{H_2^+} \cos^2 \gamma_2 &=&  {({\cal B}_C^2)}_{33}\,.
\end{eqnarray}
\end{subequations}
These equations in conjunction with Eq.~(\ref{e:BC2}) will enable us to solve for $\lambda_{7},\lambda_8$, and $\lambda_9$ as given below:
\begin{subequations} \label{e:lam789}
	\begin{eqnarray}
 \lambda_7  &=& 
 \frac{\left(m_{H_1^+}^2 - m_{H_2^+}^2\right)}{2 v^2} \left[(-3 + c_{2\beta_2})\frac{c_{2\gamma_2}}{c_{\beta_2}^2} + \frac{4\tan \beta_2}{\tan 2\beta_1} \frac{ s_{2\gamma_2}}{c_{\beta_2}}\right] -
  \frac{\left(m_{H_1^+}^2 + m_{H_2^+}^2\right)}{v^2} \nonumber \\
  && -\lambda_{10}\frac{\tan\beta_2}{ s_{\beta_1} } -\lambda_{11} \frac{\tan\beta_2}{c_{\beta_1}}  
   +  \frac{2 m_{12}^2 }{ v^2 s_{\beta_1} c_{\beta_1} c_{\beta_2}^2 }\,, \\
\lambda_8  &=&\frac{m_{H_1^+}^2}{v^2}\left(-2s^2_{\gamma_2}+\tan\beta_1 \frac{s_{2\gamma_2} }{s_{\beta_2} }\right)
-\frac{m_{H_2^+}^2}{v^2}\left(2c^2_{\gamma_2}+\tan\beta_1  \frac{ s_{2\gamma_2}}{s_{\beta_2}}\right) \nonumber \\
&&
- \lambda_{10} s_{\beta_1} \cot\beta_2 -\lambda_{12} \tan\beta_1 +  \frac{2 m_{13}^2 }{ v^2 c_{\beta_1} s_{\beta_2} c_{\beta_2}}\,,  \\
\lambda_9  &=& -\frac{m_{H_1^+}^2}{v^2}\left(2s^2_{\gamma_2}+\cot\beta_1 \frac{s_{2\gamma_2} }{ s_{\beta_2}}\right)
+\frac{m_{H_2^+}^2}{v^2}\left(-2c^2_{\gamma_2}+\cot\beta_1  \frac{s_{2\gamma_2} }{s_{\beta_2}}\right) \nonumber \\
&&
-\lambda_{11} c_{\beta_1} \cot\beta_2 -\lambda_{12} \cot\beta_1 +  \frac{2 m_{23}^2 }{ v^2 s_{\beta_1} s_{\beta_2} c_{\beta_2}}\,. 
	\end{eqnarray}
\end{subequations}
where, the other three couplings $(\lambda_{10},\lambda_{11 }~\& ~\lambda_{12})$ can be replaced using Eq.~(\ref{e:lam12}). 


\subsection{CP-even scalar sector}
The mass terms in the neutral scalar sector can be extracted from the potential as,
	\begin{subequations} 
	\begin{eqnarray}\label{e:neutralmassmat}
	V_{S}^{\rm mass} =\begin{pmatrix}
	h_1 & h_2 & h_3\\
	\end{pmatrix} \frac{{\cal M}_S^2}{2} \begin{pmatrix}
	h_1\\  h_2\\ h_3\\
	\end{pmatrix} \,,
	\end{eqnarray}
where, ${\cal M}_S^2$ is the $3\times3$ symmetric
mass matrix whose elements are given by,
	\begin{eqnarray}
	{({\cal M}_S^2)}_{11} &=& 2 v_1^2 \lambda_1 - \frac{v_2 v_3 \left(v_2 \lambda_{11} + v_3 \lambda_{12}\right)}{2 v_1} + m_{12}^2\frac{v_2}{v_1} + m_{13}^2\frac{v_3}{v_1} \,, \\
	{({\cal M}_S^2)}_{12} &=&  v_1\left(v_2 (\lambda_{7}+ \lambda_4)+ v_3 \lambda_{10}\right) + \frac{v_3}{2} \left(2 v_2 \lambda_{11} + v_3 \lambda_{12}\right) - m_{12}^2 \,, \\
	{({\cal M}_S^2)}_{13} &=&  v_1\left(v_3 (\lambda_8+ \lambda_5)+ v_2 \lambda_{10}\right)+ \frac{v_2}{2} \left( v_2 \lambda_{11} +2 v_3 \lambda_{12}\right) - m_{13}^2 \,, \\
	{({\cal M}_S^2)}_{22} &=& 2 v_2^2 \lambda_2 - \frac{v_1 v_3 \left(v_1 \lambda_{10} + v_3 \lambda_{12}\right)}{2 v_2} + m_{12}^2\frac{v_1}{v_2} + m_{23}^2\frac{v_3}{v_2}\,, \\ 
	{({\cal M}_S^2)}_{23} &=& v_3\left( v_2 (\lambda_{6}+ \lambda_9)+ v_1 \lambda_{12}\right) + \frac{v_1}{2} \left(2 v_2 \lambda_{11} + v_1 \lambda_{10}\right) - m_{23}^2 \,, \\
	{({\cal M}_S^2)}_{33} &=& 2 v_3^2 \lambda_3 - \frac{v_1 v_2 \left(v_1 \lambda_{10} + v_2 \lambda_{11}\right)}{2 v_3} + m_{13}^2\frac{v_1}{v_3} + m_{23}^2\frac{v_2}{v_3} \,.
	\end{eqnarray}\label{e:mselement}
\end{subequations}
 This mass matrix should be diagonalized via the following orthogonal transformation
 \begin{eqnarray}\label{e:msdiag}
{\cal O}_\alpha \cdot {\cal M}_{S}^2 \cdot {\cal O}_\alpha^T &\equiv& \begin{pmatrix}
 m_h^2 & 0 & 0 \\
 0& m_{H_1}^2 & 0 \\
 0 & 0 & m_{H_2}^2 \\
 \end{pmatrix}  \,,
\end{eqnarray}
where, ${\cal O}_\alpha$ has already been defined in Eq.~(\ref{e:Oa}).
Inverting the above Eq.~(\ref{e:msdiag}), we get,
\begin{eqnarray}\label{e:MS}
  {\cal M}_{S}^2  &\equiv& {\cal O}_\alpha^T  \cdot \begin{pmatrix}
m_h^2 & 0 & 0 \\
0& m_{H_1}^2 & 0 \\
0 & 0 & m_{H_2}^2 \\
\end{pmatrix} \cdot {\cal O}_\alpha\,,
\end{eqnarray}
which enables us to solve for the remaining six lambdas
as follows:
\begin{subequations}
	\label{e:lam1to6}
	\begin{eqnarray}
	\lambda_1  &=& \frac{m_h^2}{2 v^2} \frac{c^2_{\alpha_1} c^2_{\alpha_2} }{c^2_{\beta_1} c^2_{\beta_2}}
	+ \frac{m_{H_1}^2}{2 v^2 c^2_{\beta_1} c^2_{\beta_2} }  \left( c_{\alpha_1} s_{\alpha_2}  s_{\alpha_3} + s_{\alpha_1} c_{\alpha_3}  \right)^2 
	+ \frac{m_{H_2}^2}{2 v^2 c^2_{\beta_1} c^2_{\beta_2}} \left(c_{\alpha_1}   s_{\alpha_2}  c_{\alpha_3} - s_{\alpha_1} s_{\alpha_3}  \right)^2 \nonumber \\
	&&+\frac{\tan\beta_1  \tan \beta_2}{4c^2_{\beta_1}} \left( \lambda_{11} s_{\beta_1} + \lambda_{12} \tan\beta_2  \right) - \frac{m_{12}^2}{2 v^2}  \frac{\tan\beta_1}{ c^2_{\beta_1} c^2_{\beta_2}}
	-  \frac{m_{13}^2}{2 v^2}  \frac{\tan\beta_2}{ c^3_{\beta_1} c^2_{\beta_2}} \,, \\
	\lambda_2  &=& \frac{m_h^2}{2 v^2}\frac{s^2_{\alpha_1} c^2_{\alpha_2}  } {s^2_{\beta_1}c^2_{\beta_2}  }
	+\frac{m_{H_1}^2}{2 v^2 s^2_{\beta_1} c^2_{\beta_2}} \left(c_{\alpha_1} c_{\alpha_3} - s_{\alpha_1}s_{\alpha_2} s_{\alpha_3} \right)^2
	+\frac{m_{H_2}^2}{2 v^2 s^2_{\beta_1} c^2_{\beta_2}} \left(c_{\alpha_1} s_{\alpha_3} + s_{\alpha_1}s_{\alpha_2} c_{\alpha_3} \right)^2\nonumber \\
	&&+\frac{ \tan\beta_2}{4s^2_{\beta_1}\tan \beta_1 } \left( \lambda_{10} c_{\beta_1}+ \lambda_{12} \tan\beta_2\right) - \frac{m_{12}^2}{2 v^2}  \frac{\cot\beta_1}{ s^2_{\beta_1} c^2_{\beta_2}}
	-  \frac{m_{23}^2}{2 v^2}  \frac{\tan\beta_2}{ s^3_{\beta_1} c^2_{\beta_2}} \,, \\
	\lambda_3  &=& \frac{m_h^2}{2v^2} \frac{s^2_{\alpha_2}}{s^2_{\beta_2}}
	+\frac{m_{H_1}^2 c^2_{\alpha_2} s^2_{\alpha_3}}{2v^2 s^2_{\beta_2}}  
	+\frac{m_{H_2}^2 c^2_{\alpha_2} c^2_{\alpha_3}}{2v^2 s^2_{\beta_2}} 
	+\frac{s_{2\beta_1}  }{8\tan ^3\beta_2}  \left( \lambda_{10}c_{\beta_1} + \lambda_{11}s_{\beta_1}  \right) \nonumber \\ 
	&& - \frac{m_{13}^2}{2 v^2}  \frac{c_{\beta_1}}{ \tan \beta_2  s^2_{\beta_2}}
	-  \frac{m_{23}^2}{2 v^2}  \frac{s_{\beta_1}}{\tan \beta_2  s^2_{\beta_2} }
	 \,,  \\
	\lambda_4  &=& 
	\frac{1}{4v^2 s_{2\beta_1} c^2_{\beta_2}}\left[\left(m_{H_1}^2-m_{H_2}^2\right) \left\{(-3 + c_{2\alpha_2})s_{2\alpha_1}c_{2\alpha_3} - 4 c_{2\alpha_1} s_{\alpha_2} s_{2\alpha_3}\right\}
	-2\left(m_{H_1}^2+m_{H_2}^2\right)s_{2\alpha_1} c^2_{\alpha_2}  \right] \nonumber \\
	&&+\frac{m_h^2}{v^2}\frac{  s_{2\alpha_1}c^2_{\alpha_2}}{s_{2\beta_1} c^2_{\beta_2}} 
	- \frac{\tan\beta_2}{ s_{2\beta_1}}\left(2\lambda_{10}c_{\beta_1} + 2\lambda_{11}s_{\beta_1} + \lambda_{12} \tan\beta_2 \right)-\lambda_{7}  + \frac{m_{12}^2}{v^2}  \frac{1}{ s_{\beta_1} c_{\beta_1} c^2_{\beta_2}} \,, \\
	\lambda_5 &=& \frac{m_h^2}{ v^2 } \frac{c_{\alpha_1} s_{2\alpha_2}}{c_{\beta_1} s_{2\beta_2}}
	-\frac{m_{H_1}^2}{v^2c_{\beta_1} s_{2\beta_2}} \left(c_{\alpha_1} s_{2\alpha_2} s^2_{\alpha_3}  + s_{\alpha_1}  c_{\alpha_2}s_{2\alpha_3} \right)  
    +\frac{m_{H_2}^2}{v^2c_{\beta_1} s_{2\beta_2}} \left( s_{\alpha_1} c_{\alpha_2}  s_{2\alpha_3}-c_{\alpha_1} s_{2\alpha_2} c^2_{\alpha_3}\right)  \nonumber \\
	&&- \frac{s_{\beta_1}}{2\tan\beta_2} \left( 2\lambda_{10} + \lambda_{11}\tan\beta_1\right) -\lambda_{12} \tan\beta_1- \lambda_8 + \frac{m_{13}^2}{v^2}  \frac{1}{ c_{\beta_1} s_{\beta_2} c_{\beta_2}}  \,, \\
	\lambda_6  &=&  \frac{ m_h^2 }{ v^2}\frac{s_{\alpha_1}s_{2\alpha_2} } {s_{\beta_1} s_{2\beta_2}} 
	+ \frac{m_{H_1}^2 }{v^2}\frac{c_{\alpha_2}}{s_{\beta_1} s_{2\beta_2}} \left(-2s_{\alpha_1} s_{\alpha_2} s^2_{\alpha_3}  + c_{\alpha_1} s_{2\alpha_3} \right)  
	- \frac{m_{H_2}^2}{v^2} \frac{c_{\alpha_2}} {s_{\beta_1}s_{2\beta_2} } \left(2s_{\alpha_1} s_{\alpha_2} c^2_{\alpha_3}  + c_{\alpha_1} s_{2\alpha_3} \right)\nonumber \\
	&& - \frac{c_{\beta_1}}{2\tan\beta_2} \left(\lambda_{10} \cot\beta_1  +2 \lambda_{11}\right) -\lambda_{12} \cot\beta_1-\lambda_{9} + \frac{m_{23}^2}{ v^2}  \frac{1}{ s_{\beta_1} s_{\beta_2} c_{\beta_2}}  \,.
	\end{eqnarray} 
\end{subequations}
%





\section{Flavor observables in the Z3HDM}\label{app:flavorObs}

\subsection{Computing $ b \to s \gamma$}
\begin{figure}[htbp!]
\begin{minipage}{.5\textwidth}
  \centering
  \includegraphics[width=.7\linewidth]{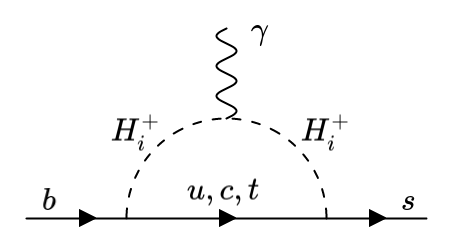}
\end{minipage}%
\begin{minipage}{.5\textwidth}
  \centering
  \includegraphics[width=.7\linewidth]{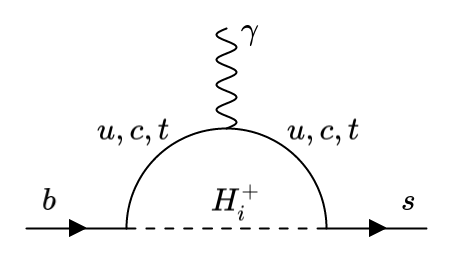}
\end{minipage}
\caption{\small NP contributions to $b \to s \gamma$ in the Z3HDM. $H_i^\pm$ stands for both charged scalars ($i=1,2$).}
\label{fig:b2sg}
\end{figure}

The nonstandard contributions to the one-loop $b\to s\gamma$ amplitude in our Z3HDM scenario
are shown in Fig.~\ref{fig:b2sg}. Since the one-loop contributions come from the charged scalar
only, the NP amplitudes will depend only on the parameters $\tan\beta_1$, $\tan\beta_2$,
$m_{H_1^+}$, $m_{H_2^+}$ and $\gamma_2$.
To find the amplitudes, we simply extend the analysis of a NFC 2HDM \cite{Deschamps:2009rh, Das:2015gyv} for a scenario with two different $H^+$.
Following Ref.~\cite{Gambino:2001ew}, the branching ratio for $b \to s \gamma$ is controlled by the $C_{7L}^\text{eff}$ and $C_{7R}^\text{eff}$ Wilson coefficients:
\begin{equation}
\frac{\text{Br}\left( b \to s \gamma \right)}{\text{Br}\left( b \to c e \overline{\nu} \right)} = \frac{6 \alpha}{\pi B} {\bigg\lvert \frac{V_{ts}^* V_{tb}}{V_{cb}} \bigg\rvert}^2\left[ {\big\lvert C_{7L}^\text{eff} \big\rvert}^2+{\big\lvert C_{7R}^\text{eff} \big\rvert}^2\right],
\end{equation}
where the normalization by $\text{Br}\left( b \to c e \overline{\nu}\right)$ helps canceling some of the hadronic uncertainties. The effective Wilson coefficients read
\begin{subequations}
	\label{eq:effWilson}
\begin{eqnarray}
	C_{7L}^\text{eff} &=& \eta^{16/23} C_{7L} + \frac{8}{3}\left( \eta^{14/23}-\eta^{16/23} \right) C_{8L} + \sum_{i=1}^8 h_i  \eta^{a_i} \,,  \label{eq:C7Leff}\\
	C_{7R}^\text{eff} &=& \eta^{16/23} C_{7R} + \frac{8}{3}\left( \eta^{14/23}-\eta^{16/23} \right) C_{8R} ,\label{eq:C7Reff}
\end{eqnarray}
where, as in the usual analysis of 2HDMs \cite{Gambino:2001ew}, the leading $\log$ QCD corrections in the SM are described by
\begin{eqnarray}
&&a_i = \begin{pmatrix} \frac{14}{23}, & \frac{16}{23}, & \frac{6}{23}, & -\frac{2}{23}, & 0.4086, & -0.4230, & -0.8994, & 0.1456 \end{pmatrix}, \\
&&h_i =\begin{pmatrix} \frac{626126}{272277}, & -\frac{56281}{51730}, & -\frac{3}{7}, & -\frac{1}{14}, & -0.6494, & -0.0380, & -0.0186, & -0.0057   \end{pmatrix},
\end{eqnarray}
\end{subequations}
and $\eta = \alpha_s(M_Z)/\alpha_s(\mu)$, where $\mu$ is the QCD renormalization scale, $\mu \approx 221$ MeV.
Taking into account the absence of tree-level FCNCs, the coefficients in \Eqs{eq:C7Leff}{eq:C7Reff} can be recast as 
\begin{subequations}
\begin{eqnarray}
&&C_{7L} = A_\gamma^\text{SM} + A^+_{\gamma L}, \qquad C_{7R} = \frac{m_s}{m_b} A_\gamma^\text{SM} + A^+_{\gamma R}, \\
&&C_{8L} = A_g^\text{SM} + A^+_{g L}, \qquad C_{8R} =  \frac{m_s}{m_b} A_g^\text{SM} + A^+_{g R}, 
\end{eqnarray}
\end{subequations}
where the $A^+$ terms correspond to our NP (charged-Higgs) contributions. These contributions can be further broken down into
\begin{subequations}\label{eq:NPcontrs}
\begin{eqnarray}
&&A^+ _{\gamma L, R} = \frac{1}{V_{ts}^* V_{tb}} \sum_{q=u,c,t} V^*_{qs}V_{qb} \left[ C_{1L, R} (y_q) + \frac{2}{3} C_{2L, R}(y_q)\right],\\
&&A^+ _{g L, R} = \frac{1}{V_{ts}^* V_{tb}} \sum_{q=u,c,t} V^*_{qs}V_{qb}  C_{2L, R}(y_q),
\end{eqnarray}
\end{subequations}
with $y_q=m_q^2/M^2_{H^+}$ and 
\begin{subequations}\label{eq:NPcoefs}
\begin{eqnarray}
&&C_{1L, R}(y_q) = \frac{y_q}{4} \left( \bigg[ \overline{\mathcal{F}_2}(y_q)- \overline{\mathcal{F}_1}(y_q) \bigg] \left(\frac{m_{s, b}^2}{m_q^2} Y^2 + X^2 \right) + 2 XY \bigg[ \overline{\mathcal{F}_1}(y_q)- \overline{\mathcal{F}_0}(y_q) \bigg] \right), \\
&&C_{2L, R}(y_q) = \frac{y_q}{4} \left( \bigg[ \mathcal{F}_2(y_q)- \mathcal{F}_1(y_q) \bigg] \left(\frac{m_{s, b}^2}{m_q^2} Y^2 + X^2 \right) - 2 XY \mathcal{F}_1(y_q) \right),
\end{eqnarray}
\end{subequations}
in which $X$ and $Y$ are the charged-Higgs coupling to left- and right-handed quarks, respectively, and the loop functions are given by 
\begin{subequations}\label{eq:hypergeofn}
\begin{eqnarray}
&&\mathcal{F}_k (t) = \int_0^1 dx \frac{(1-x)^k}{x+(1-x)t} =  \frac{1}{\left(k+1\right) t}\,\,\, \pFq{2}{1}{1,1}{}{k+2; \frac{t-1}{t}},\\
&&\overline{\mathcal{F}}_k (t) = \int_0^1 dx \frac{x^k}{x+(1-x)t} =   \frac{1}{\left(k+1\right) t}\,\,\,  \pFq{2}{1}{1,k+1}{}{k+2; \frac{t-1}{t}},
\end{eqnarray}
\end{subequations}
where $\pFq{p}{q}{a,b}{}{c;d}$ is the Hypergeometric Function. Finally, the SM amplitude is given by (keeping only the top contribution)
\begin{subequations}
\begin{eqnarray}
 A_\gamma^\text{SM} &=& \bigg[   \frac{\left( 2- 3x_t \right)}{2}\overline{\mathcal{F}_1} (x_t) + \frac{\left(2+ x_t\right)}{2}  \overline{\mathcal{F}_2} (x_t)  + x_t \overline{\mathcal{F}}_0(x_t)  \nonumber   \\ 
&& \qquad \,\, + \frac{4}{3} \mathcal{F}_0(x_t) - \frac{\left(6- x_t\right)}{3} \mathcal{F}_1 (x_t) + \frac{\left( 2 + x_t \right)}{3} \mathcal{F}_2(x_t) \bigg] - \frac{23}{36}, \\
 A_g^\text{SM} &=&\bigg[  2 \mathcal{F}_0(x_t) - \frac{\left(6-x_t \right)}{2} \mathcal{F}_1(x_t) + \frac{\left(2+x_t\right)}{2} \mathcal{F}_2 (x_t) \bigg] - \frac{1}{3},
\end{eqnarray}
\end{subequations}
where $x_t=m^2_t/M_W^2$. So far, we have presented the analysis of the $b \to s \gamma$ processes in a 2HDM where FCNCs are absent. To extend these results to our model, we redefine Eqs.~\eqref{eq:NPcontrs} and \eqref{eq:NPcoefs} to account for both charged-Higgs contributions:
\begin{subequations}\label{eq:NPcontrsZ3HDM}
\begin{eqnarray}
&&A^+ _{\gamma L, R} = \frac{1}{V_{ts}^* V_{tb}} \sum_{q=u,c,t} V^*_{qs}V_{qb} \sum_{i=1,2}\left[ C^i_{1L, R} (y^i_q) + \frac{2}{3} C^i_{2L, R}(y^i_q)\right],\\
&&A^+ _{g L, R} = \frac{1}{V_{ts}^* V_{tb}} \sum_{q=u,c,t} V^*_{qs}V_{qb} \sum_{i=1,2} C^i_{2L, R}(y^i_q),
\end{eqnarray}
\end{subequations}
where now $y^i_q=m_q^2/M^2_{H^+_i}$, and 
\begin{subequations}\label{eq:NPcoefsZ3HDM}
\begin{eqnarray}
&&C^i_{1L, R}(y_q) = \frac{y_q}{4} \left( \bigg[ \overline{\mathcal{F}_2}(y_q)- \overline{\mathcal{F}_1}(y_q) \bigg] \left(\frac{m_{s, b}^2}{m_q^2} Y_i^2 + X_i^2 \right) + 2 X_i Y_i  \bigg[ \overline{\mathcal{F}_1}(y_q)- \overline{\mathcal{F}_0}(y_q) \bigg] \right), \\
&&C^i_{2L, R}(y_q) = \frac{y_q}{4} \left( \bigg[ \mathcal{F}_2(y_q)- \mathcal{F}_1(y_q) \bigg] \left(\frac{m_{s, b}^2}{m_q^2} Y_i^2 + X_i^2 \right) - 2 X_i Y_i \mathcal{F}_1(y_q) \right),
\end{eqnarray}
\end{subequations}
where we can see the $X_i$ and $Y_i$ couplings now carry an index, denoting the $H_1^+$ and $H_2^+$ chiral ($P_L$ and $P_R$) couplings to quarks. In the present model, these couplings can be extracted from Eqs.~\eqref{yukla5} and \eqref{yukla6}:
\begin{subequations}
	\begin{eqnarray}\label{eq:vertices}
	&&X_1 = - \cot\beta_2 \sin \gamma_2, \\
	&&Y_1= - \tan \beta_2 \left( \frac{\cot\beta_1 \cos\gamma_2}{\sin\beta_2} + \sin\gamma_2 \right), \\ 
	&&X_2 = \cot \beta_2 \cos \gamma_2, \\
	&&Y_2 = - \tan \beta_2 \left( \frac{\cot\beta_1 \sin\gamma_2}{\sin\beta_2} - \cos\gamma_2 \right).
	\end{eqnarray}
\end{subequations}
We now have  all the relevant information needed to compute the $b \to s \gamma$ branching ratio in our model. As advertised, the only dependencies on the BSM degrees of freedom is through $\tan\beta_1$, $\tan\beta_2$, $\gamma_2$, which control the couplings, the charged-Higgs masses, $m_{H^+_1}$ and $m_{H^+_2}$, which will affect the loop functions. Finally, the SM prediction for the $b \to s \gamma$ branching ratio can be found in ref~\cite{Misiak:2020vlo}, and the experimental values in \cite{Zyla:2020zbs}. 

\subsection{Neutral Meson Mixing: $ \Delta M_{Bq}$}
\begin{figure}[h]
	\centering
\begin{minipage}{.3\textwidth}
  \centering
  \includegraphics[width=.9\linewidth]{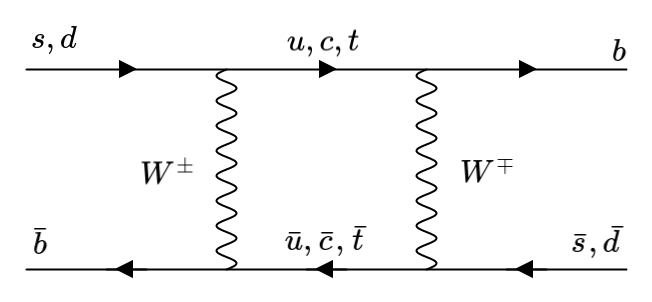}
\end{minipage}%
\begin{minipage}{.3\textwidth}
  \centering
  \includegraphics[width=.9\linewidth]{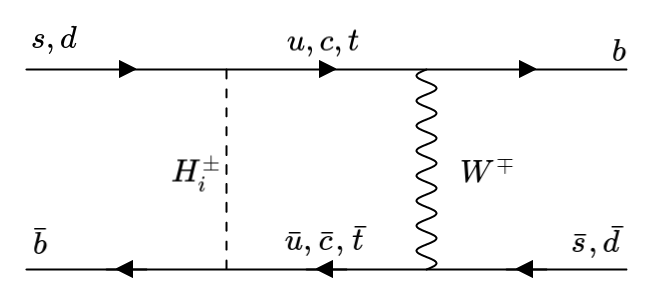}
\end{minipage}
\begin{minipage}{.3\textwidth}
  \centering
  \includegraphics[width=.9\linewidth]{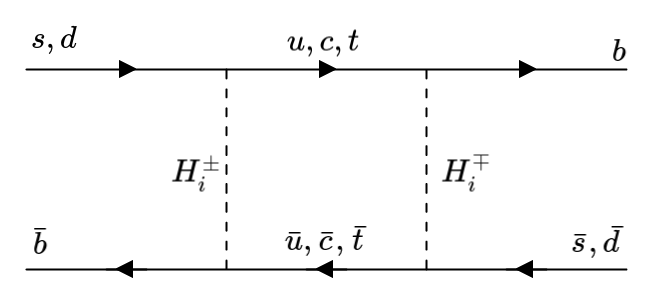}
\end{minipage}
\caption{\small Contributions to $\Delta M_{B_q}$ in the Z3HDM. $H_i^\pm$ stands for both charged scalars ($i=1,2$). The first box diagram corresponds to the SM amplitude. The diagrams with interchanged internal lines are not shown explicitly.}
\label{fig:DeltaMdiags}
\end{figure}

A very restrictive aspect of BSM models comes from neutral meson oscillations. These processes, for models without tree-level FCNCs, are forbidden at tree-level, but may have sizable one-loop contributions. The left panel in Fig.~\ref{fig:DeltaMdiags} represents the SM contribution for
such processes, whereas the other two diagrams represent the additional contributions in our
Z3HDM scenario.
To obtain some qualitative intuitions we write the effective $\Delta F=2$ Lagrangian as:
\begin{equation}\label{eq:LeffDeltaM}
\mathcal{L}^{\Delta F=2}_\text{eff} =  \frac{G_F^2 M_W^2}{16 \pi^2} \!\!\!\!\!\!\!\!
\sum_{\substack{
		a,b=u,c,t \\
                  i,j = H_1^\pm, H_2^\pm}
                  }\!\!\!\!\!\!\!\!
\lambda_a \lambda_b \,\, \omega_a \omega_b \bigg( \frac{S(y_a, y_b)}{4} + X_{ia} X_{ib} \bigg[ I_1(y_a, y_b, y_{i}) + X_{ja} X_{jb} I_2 (y_a, y_b, y_i, y_j)  \bigg] \bigg) O_F.
\end{equation}
The SM contribution is encoded in $S(y_a, y_b)$, normalized by a factor of $4$ to account for the summation on the charged Higgs. The $I_1(y_a, y_b, y_i)$ contributions are due to the mixed $W^\pm - H_i^\pm$ boxes, and $I_2(y_a, y_b, y_i, y_j)$ are the $H_i^\pm - H_j^\pm$ boxes in Fig.~\ref{fig:DeltaMdiags}. The above expression is valid in the zero external momenta approximation, where the down-type quark masses are taken to be zero. We use $X_{ia}$ to denote the coupling between the charged-Higgs $H_i^\pm$ and the up-quark $a$, which, as seen in Eq.~\eqref{eq:vertices}, are flavor universal, {\it i.e.}, $X_{1a} = X_1 = -\cot\beta_2 \sin\gamma_2$ and $X_{2a}= X_2 = \cot\beta_2 \cos\gamma_2$ for $H_1^\pm$ and $H_2^\pm$, respectively.
The quantities $y_a$ and $y_i$ stand for the ratios $m^2_a/M^2_W$ and $m^2_{H^+_i}/M_W^2$
respectively.
The specificities of the neutral meson under consideration are contained 
 in the CKM elements $\lambda_a$, and the dimension-6 operators $O_F$. For a generic meson $P= \left(\overline{q_1}, q_2 \right)$, these are defined as
\begin{equation}
\lambda_a = \left( V^*_{a \,q_2} V_{a \,q_1} \right), \qquad O_F = \left(\overline{q}_1 \gamma^\mu P_L q_2 \right)^2.
\end{equation}
Finally, the loop functions are given by
\begin{subequations}
	\begin{eqnarray}
	&&f(x) = \frac{\left(x^2-8x+4\right)\ln x +3(x-1)}{(x-1)^2}, \qquad S(y_a, y_b) = \frac{f(y_a, y_b)}{y_a- y_b}, \\
	&&g(x,y,z) = \frac{x(x-4)\ln x}{(x-1)(x-y)(x-z)}, \quad I_1(y_a, y_b, y_i) =  g(y_a, y_b, y_i)+g(y_b, y_i, y_a)+g(y_i, y_a, y_b), \\
	&& h(x,y,w,z) = \frac{ x^2 \ln x}{(x-y)(x-w)(x-z)}, \\ 
	&& I_2(y_a, y_b, y_i, y_j) =  h(y_a, y_b, y_i, y_j)+h(y_b, y_a, y_i, y_j)  +h(y_i, y_a, y_b, y_j)+h(y_j, y_a, y_b, y_i) .
	\end{eqnarray}
\end{subequations}
The limiting cases where, for instance, the same Higgs runs in the $I_2$ box diagram should be carefully dealt with, as the loop functions are only apparently divergent for $x_i=x_j$, but indeed have a well-defined limit. 

Finally, we can obtain $\Delta M_P$ from the effective Lagrangian,
\begin{subequations}
	\begin{eqnarray}
	&&\Delta M_P = 2 \lvert M_{12}^P \rvert, \qquad M_{12}^P = -\frac{1}{2 M_P} \big< P^0 \big\lvert \mathcal{L}_\text{eff}^{\Delta F=2} \big\rvert \overline{P}^0 \big>, \\
	&&\big< P^0 \big\lvert O_F^P \big \rvert \overline{P}^0 \big> = \frac{2}{3} f_P^2 M_P^2 B_P,
	\end{eqnarray}
\end{subequations}
where $M_P$ is the meson mass, $f_P$ its decay constant, and $B_P$ is its bag parameter. The 2HDM limit (with no tree-level FCNCs) of Eq.~\eqref{eq:LeffDeltaM} can be easily extracted, taking some care on the symmetry factors. As in the $b \to s \gamma$ computations, it would be possible to parametrize these results to match numerical results with higher-order corrections. The experimental values which will determine the experimentally allowed region are taken from \cite{Zyla:2020zbs}, whereas the relevant hadronic parameters can be found in \cite{Aoki:2019cca}.


\bibliographystyle{JHEP}
\bibliography{Article.bib}

\end{document}